\RequirePackage{amsmath}
\documentclass[12pt]{iopart}

\usepackage{graphicx}
\usepackage{cite}
\usepackage{amsfonts}
\usepackage{bm,bbold,bbm}
\usepackage[urlcolor=blue]{hyperref} 
\hypersetup{colorlinks=true,allcolors=blue}
\usepackage{amsmath,amscd,amsbsy,amssymb,latexsym,url,amsthm}
\usepackage{xcolor}

\usepackage{placeins} 

\renewcommand{\mp}{\mathbbm{P}}

\renewcommand{\aa}{\alpha}

\newcommand{\ind}[1]{\mathbbm{1}_{#1}}

\newcounter{appdxsect}
\def\alphainsection{0}

\let\oldsection=\section
\def\section{%
  \ifnum\alphainsection=1%
    \addtocounter{appdxsect}{1}
  \fi%
\oldsection}%

\renewcommand\thesection{%
  \ifnum\alphainsection=1%
    \Alph{appdxsect}%
  \else%
    \arabic{section}%
  \fi%
}%

\newenvironment{appdxsection}{%
  \ifnum\alphainsection=1%
    \errhelp={Let other blocks end at the beginning of the next block.}
    \errmessage{Nested Alpha section not allowed}
  \fi%
  \setcounter{appdxsect}{0}
  \def\alphainsection{1}
}{%
  \setcounter{appdxsect}{0}
  \def\alphainsection{0}
}%

\begin{document}


\title[Towards Markov-State Holography]{Towards Markov-State Holography}

\author{Xizhu Zhao$^1$\orcid{0000-0002-3400-9586}, Dmitrii E. Makarov$^{2,3}$\orcid{0000-0002-8421-1846} and Alja\v{z} Godec$^{1,*}$\orcid{0000-0003-1888-6666}}

\address{$^1$ Mathematical bioPhysics Group, Max Planck Institute for Multidisciplinary Sciences, 37077 G\"ottingen, Germany}

\address{$^2$ Department of Chemistry, University of Texas at Austin, Austin, Texas 78712, USA}

\address{$^3$ Oden Institute for Computational Engineering and Sciences, University of Texas at Austin, Austin, Texas 78712, USA}

\address{$^*$ Author to whom any correspondence should be addressed.}

\ead{agodec@mpinat.mpg.de}


\begin{abstract}
    Experiments, in particular on biological systems, typically probe lower-dimensional observables which are projections of high-dimensional dynamics. In order to infer consistent models capturing the relevant dynamics of the system, it is important to detect and account for the memory in the dynamics. 
    We develop a method to infer the presence of hidden states and transition pathways based on observable transition probabilities conditioned on history sequences of visited states for projected (i.e.~observed) dynamics of Markov processes. 
    Histograms conditioned on histories reveal information on the transition probabilities of hidden paths \emph{locally} between any specific pair of observed states. The convergence rate of these histograms towards a stationary distribution provides \emph{a local quantification of the duration of memory}, which reflects how distinct microscopic paths projecting onto the same observed transition decorrelate in path space. This motivates the notion of ``weak Markov order'' and provides insight about the hidden topology of microscopic paths in a holography-like fashion. The method can be used to test for the local Markov property of observables. The information extracted is also helpful in inferring relevant hidden transitions which are not captured by a Markov-state model. 
\end{abstract}

\section{Introduction}
Many processes in complex, especially living, systems are described as Markov diffusion on an energy landscape \cite{Papoian,Wales_2018,landscapes_24}, or---at the longest timescales---as continuous-time Markov jump processes \cite{Chu_2017,Pande,Shafiekhani_2021}.
Experiments, however, typically probe low-dimensional observables, which are projections of high-dimensional dynamics that are ``blind'' to hidden degrees of freedom \cite{Plotkin1998,Swope2004,Min2005,Noe2008,Sangha2009,Stigler2011,Narinder2018, Pyo2019, Vollmar2024,Baiesi_GLE,Baiesi_TUR,Baiesi_2024}. 
The observed dynamics therefore displays memory (i.e.~is non-Markov) whenever the hidden degrees of freedom do not relax much faster than the observed ones
\cite{Zwanzig1960,Mori1965,Haken1975,Zwanzig2001,Makarov2013,Avdoshenko2017,Ozmaian2019,Lapolla2019,Ayaz2022,Tanja_PR,Ginot2022,Roland_PRL,Lavacchi_2024,Dalton_2025,Lindner,Kay_1,Kay_2}. In such a non-Markov process, the state changes depend not only on the current state but also on the past.

The typical way to account for memory in continuous-space systems is via generalized Langevin equations, which require the (highly non-trivial) sampling and parameterization of a potential of mean force and a memory kernel \cite{Tanja_PR,Tanja_JCP,Tanja_3,Ginot2022,Baiesi_GLE,Roland_PRL,Lavacchi_2024,Dalton_2025}.
A widely adopted strategy to account for hidden states in discrete-state systems are Hidden-Markov Models (HMMs) \cite{HMM_2017}. However, HMMs inherently rely on explicit assumptions about the underlying model, which, especially because of limited sampling and over-fitting \cite{Hastie_2009} as well as Bayesian prior-dependence \cite{critique,Liseo_2000} give rise to hard challenges in model selection. 
``Model-free'' inference methods---in the present context, methods that make \emph{no} assumptions about the underlying topology and number of states in the network and do \emph{not} require a Markov model to even be parameterized---are therefore desired.

Several approaches have been developed to infer and quantify memory from time-series in a model-free manner
\cite{Berezhkovskii2018,Satija2020,Hartich2021,Lapolla2021,Song2023,Lindner,Zhao2024,Vollmar2024}.    
One common approach for testing for Markovianity is examining the distribution of the waiting times in each observed state \cite{Talaga2007, Cornish2008, Stigler2011, Deguenther2024a}. However, exponentially distributed waiting-times are only a necessary but \emph{not} a sufficient condition for a process on a discrete state space to be Markov. While a non-exponential waiting time distribution certainly indicates memory,
the converse is \emph{not} true; it is easy to construct non-Markov models with exponential waiting time distributions \cite{Hartich_2024_cm}. More sensitive methods exploit the statistical properties of transition-path times to infer the existence of hidden transition pathways \cite{Berezhkovskii2018,Satija2020,Hartich2021},  test for violations of the Chapman-Kolmogorov (i.e.\ semigroup) property \cite{Lapolla2021,Zhao2024,Vollmar2024}, or use Shannon's information-theoretical approach to detect memory in observed time series \cite{Song2023}. However, such approaches test for non-Markovianity of the entire observed time series, and none of these methods can detect and quantify memory \emph{locally}. 

In related studies, great efforts have been made in the context of thermodynamic inference from coarse and partial observations \cite{Mehl2012,Roldan2010,Roldan2012,Seifert2012,Esposito2012,Harunari2022,Meer2022,Meer2023,Deguenther2024,Deguenther2024a,Blom2024,Dieball_perspective,Tassilo}. In particular, the existence of the so-called ``Markovian events''---states, or transitions between states where the microscopic is fully observable---have proven to be a crucial assumption in many methods of thermodynamic inference
\cite{Meer2023,Deguenther2024,Deguenther2024a}. But since such Markovian events are local properties of the observed dynamics, it again remains elusive to reliably (and conclusively) detect them. More even than the practical need to analyze experimental data, we face the fundamental question:~\emph{Can one infer the presence of hidden parallel transitions between specific observed states locally without specifying an underlying model?} This would also provide an important step towards detecting Markovian events.

Here, we develop a method to quantify the extent of memory \emph{locally} between any pair of observed states based on the defining Markov property. 
This method allows one to infer the presence of hidden states and transition channels from observed trajectories in a \emph{holographic} and model-free manner. The holography-like inference exploits how the transition probability between pairs of observed states depends on the history.

\section{Histograms of transition probabilities conditioned on histories }
\subsection{Setup}
We consider the full dynamics to be an $m$-state
continuous-time Markov jump process on an irreducible network with a time-homogeneous generator $\mathbf{L}=\{L_{\alpha\beta}\}$, whose off-diagonal elements $L_{\alpha\beta}$ are transition rates from state $\beta$ to $\alpha$,
and diagonal elements are $L_{\beta\beta} = -\sum_{\alpha\neq\beta}
L_{\alpha\beta}$.  
We assume that some microscopic states are projected onto, i.e. lumped into, the same observable state. 
The microscopic states lumped into the same observable state are not distinguishable in the observed dynamics, which typically leads to memory
\cite{Lapolla2019,Godec2023,Hartich2023,Vollmar2024,Zhao2024,Dieball_perspective,Tassilo}. In this work, we only consider the sequence of visited states and discard the waiting times between transitions, as shown in Figure~\ref{fig1}(b). In other words, we will only consider memory, i.e.\ non-Markovianity, in the observed sequences of states. Waiting times may be analyzed in addition; if these are non-exponentially distributed this implies memory. 
 
For any \emph{observable} transition between different observable states $j\rightarrow i$ with a preceding history sequence of length $k$, we denote the observed sequence as
$\{s_{t}\}_{1\leqslant t\leqslant k+2} = s_1s_2\cdots s_k ji$, and
define the history sequence as $S_k\equiv \{s_{t}\}_{1\leqslant t\leqslant k} = s_1s_2\cdots s_k$. Note that the states visited in any pair of consecutive steps are distinct, i.e. $\forall\, t,\, s_{t}\neq s_{t+1}$.
A corresponding microscopic (fully resolved) sequence is denoted by $\{\sigma_{\tau}\}_{1\leqslant \tau\leqslant k'} = \sigma_1 \sigma_2 \cdots \sigma_{k'}$. 
In general, the length of a microscopic sequence corresponding to $\{s_{t}\}_{1\leqslant t\leqslant k+2}$ can take any value satisfying $k'\geqslant k+2$, as some microscopic transitions may not change observable states and thus are not seen in the observed time series. Specifically, for the network in Figure~\ref{fig1}(a), $k'=k+2$ because there are no connections between microscopic states in the same observable state. 
The microscopic sequence $\{\sigma_{\tau}\}_{1\leqslant \tau\leqslant k'}$ is a discrete-time Markov chain with the transition matrix $\mathbf{\Phi}=\{\Phi_{\alpha\beta}\}$ given by the splitting probabilities of the continuous-time Markov jump process, with $\Phi_{\alpha\beta} = -L_{\alpha\beta}/L_{\beta\beta}\,(\alpha\neq\beta)$, and the diagonal elements are given by $\Phi_{\alpha\alpha}=0$ \cite{Hartich2021,Blom2024}. 

\begin{figure}[h]
    \centering
    \includegraphics[width=0.5\textwidth]{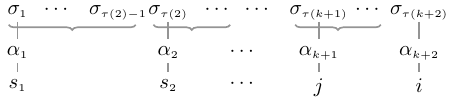}
    \caption{A schematic illustrating the definition of a microscopic full sequence $\{\sigma_{\tau}\}_{1\leqslant \tau\leqslant k'}$ (above), a reduced sequence $\{\aa_{t}\}_{1\leqslant t\leqslant k+2}$ (middle), and the observable sequence $\{s_{t}\}_{1\leqslant t\leqslant k+2}$ (below). }
    \label{fig_Appx_1}
\end{figure}

\begin{figure}[ht]
    \centering
    \includegraphics[width=0.70\textwidth]{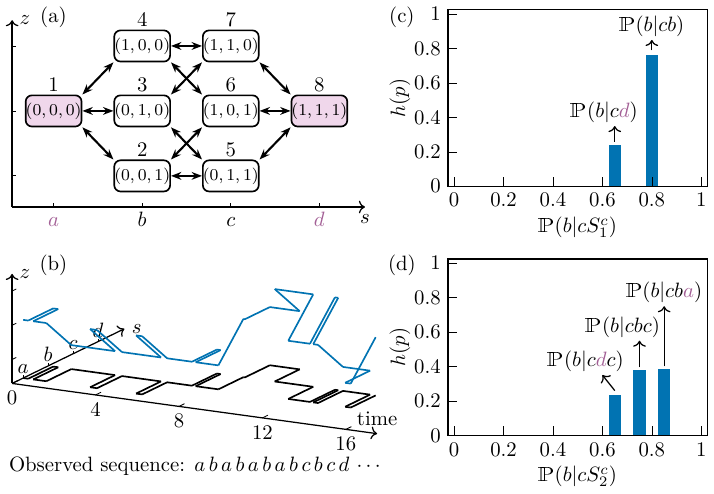}
    \caption{(a)~Schematic of the \emph{full} Markov network representing a toy protein model: the $s$-axis is the observable and the $z$-axis the hidden dimension. Microscopic states $2,3,4$ and $5,6,7$ are projected onto observable states $b$ and $c$, respectively. State $a$ and $d$ in pink are observable Markov states, as they correspond to a single microscopic state $1$ and $8$, respectively.~(b)~Example of a microscopic (blue) and the corresponding observed (black) trajectory. The latter is a projection of the microscopic trajectory onto the $s$-$t$ plane;~the corresponding observed sequence of states is shown below.~(c)~Histogram of transition probability from state $c$ to $b$ conditioned on one step in the history $\mp(b|cS^c_1)$, where $S^c_1\in\{b,d\}$. The two bars are due to $\mp(b|cd)$ and $\mp(b|cb)$.~(d)~Histogram of transition probability conditioned on two steps in the history $\mp(b|cS^c_2)$, where $S^c_2\in\{dc,bc,bd\}$. Note that the sequences in the conditional probability are visited from right to left ordered in time, e.g.\ $\mp(b|cd)$ corresponds to the sequence $d\rightarrow c\rightarrow b$.}
    \label{fig1}
\end{figure}

We further define a reduced sequence of microscopic states $\{\aa_{t}\}_{1\leqslant t\leqslant k+2}$ consisting of only the first microscopic step after every observable transition as $\aa_{t} = \sigma_{\tau(t)}$, where $\tau(1)=1$, and $\tau(t+1) = \inf\{\tau': \tau'>\tau(t),\, \ind{s_{t+1}}[\sigma_{\tau'}]=1 \}$ for any ${t\geqslant 1}$. Here, ${\mathbbm{1}_{X}}$ is the indicator function of the set $X$, taking the value $1$ if the microscopic state $\sigma_{\tau'}$ belongs to the observable state $s_{t+1}$, and taking $0$ otherwise. In other words, we construct the reduced sequence by revealing the first microscopic state in every observable step, as illustrated in Figure~\ref{fig_Appx_1}. Generally,  different $\{\aa_{t}\}_{1\leqslant t\leqslant k+2}$ can yield the  same $\{s_{t}\}_{1\leqslant t\leqslant k+2}$. The dynamics of the reduced microscopic sequence $\{\aa_{t}\}_{1\leqslant t\leqslant k+2}$ is Markov. The microscopic transition probability, i.e.~splitting probability, is given by an
effective transition matrix $\mathbf{\Omega}$, whose elements are transition probabilities $\Omega_{\aa_{t+1}\aa_{t}} = \mp(\aa_{t+1}|\aa_{t})$ (See derivation for $\mathbf{\Omega}$ in Appendix \hyperref[sec:matrix]{A}).
The probability of a reduced sequence is given by $\mp(\{\aa_{t}\}_{1\leqslant t\leqslant k}) = \prod_{i=t}^{k-1} \mp(\aa_{t+1}|\aa_{t}) \mp(\aa_1)$, where $\mp(\aa_1)$ is the steady state probability at $\aa_1$ in the reduced sequence. 

\subsection{Distribution of sequence histories}
For any pair of observed states $j$ and $i$, we denote the probability of the sequence $S_k\rightarrow j\rightarrow i$ by $\mp(ijS_k)$. 
Let $\{S^j_k\}\subset \{S_k\}$ be the set of possible (distinct) histories with length $k$ that precede $j$, which is a subset of all possible sequences with length $k$. We denote the transition probability of a generally non-Markov transition $j\to i$ conditioned on a specific history $S^j_k$ as $\mp(i|jS^j_k)$. 
We introduce a shorthand notation for the sum of probabilities of microscopic sequences with the same observable history $S^j_k$ as
\begin{align}
    \mp(\aa_{k+1}S^j_k) \equiv\!\!\!\!\!\!\sum_{\{\aa_{t}\}_{1\leqslant t\leqslant k}}\!\!\!\!\!\! \ind{S^j_k}[\{\aa_{t}\}_{1\leqslant t\leqslant k}] \mp(\{\aa_{t}\}_{1\leqslant t\leqslant k+1}),
\end{align}
where ${\mathbbm{1}_{X}}$ is the indicator function of the set $X$, taking the value $1$ if the microscopic sequence $\{\aa_{t}\}_{1\leqslant t\leqslant k}$ is projected onto the observable sequence $S^j_k$ and $0$ otherwise.~The transition probability of a generally non-Markov transition $j\to i$ conditioned on a specific history $S^j_k$ is given by
\begin{align}
     \mp(i|jS^j_k) = 
    \frac{\sum_{\aa_{k+1}}\!\! \ind{j}[\aa_{k+1}] \mp(\aa_{k+1}S^j_k) \mp(i|\aa_{k+1})}{\sum_{\aa_{k+1}} \ind{j}[\aa_{k+1}] \mp(\aa_{k+1}S^j_k)} ,
    \label{def_split_prob1}
\end{align}
where ${\mp(i|\aa_{k+1}) \equiv \sum_{\aa_{k+2}}\!\! \ind{i}[\aa_{k+2}] \mp(\aa_{k+2}|\aa_{k+1})}$ is the transition probability from microscopic state $\aa_{k+1}$ to any microscopic state in observable state $i$. 
Given a sufficiently long observed sequence, $\mp(i|jS^j_k)$ can be determined, in practice, by dividing the number of occurrences of the sequence $S^j_k\rightarrow j\rightarrow i$ by that of the sequence $S^j_k\rightarrow j$. 
Eq.~\eqref{def_split_prob1} shows that the observed transition probability from $j$ to $i$ can be expressed as a weighted average of splitting probabilities $\mp(i|\aa_{k+1})$ from a microscopic state $\aa_{k+1}$ to an observable state $i$, and the prefactors $\mp(\aa_{k+1}S^j_k)$ depend on the history. 

Different history sequences $S^j_k$ in general yield different conditional probabilities $\mp(i|jS^j_k)$: The dynamics of the observed sequence is Markov \emph{if and only if} the transition probability between any pair of states does \emph{not} depend on the history. This is true \emph{if and only if}, for any given pair $i,j$, all transition probabilities $\mp(i|jS^j_k)$ are identical,  $\mp(i|jS^j_k)=\mp(i|j)$, for all $k$.
In contrast, for non-Markov dynamics one expects different values of  $\mp(i|jS^j_k)$ for different histories  $S^j_k$. We can quantify these variations by binning different values to form a  histogram as follows:
\begin{align}
    h^\delta_k(p)\equiv \frac{1}{\mp(j)} \sum_{\{S^j_k\}}\mp(jS^j_k)\mathbbm{1}_{[p-\delta,p+\delta)}[\mp(i|jS^j_k)],
    \label{h}
\end{align}  
where ${\mathbbm{1}_{X}}$ is the indicator function of the set $X$. The conditional probability $\mp(i|jS^j_k)$ is a feature of a given $S_k$. Histories $S^j_k$ yielding $\mp(i|jS^j_k)\in [p-\delta,p+\delta)$ are classified into the same bar at $p$. Each $S^j_k$ is weighted by the sequence probability $\mp(jS^j_k)$, with a normalization factor $\mp(j)$. This yields a histogram $h^\delta_k(p)$ with $\sum_p h^\delta_k(p) =1$.

The histograms so defined can be used to infer memory systematically and, by the explicit dependence on $i$ and $j$, locally.  If there exist certain $j$ and $i$ where the histogram at some $k$ has non-zero values at locations other than $\mp(i|j)$, then the transition from $j$ to $i$ has memory, and therefore the state $j$ is non-Markov. 
We can see from Eq.~\eqref{def_split_prob1} that the positions of non-zero bars are bounded by the smallest and largest values of $\mp(i|\aa_{k+1})$.

\subsection{Convergence of histograms}
By comparing histograms for
different history length $k$, we can obtain information on the duration of \emph{local memory}.
Changes in the histograms upon increasing $k$ reflect the memory about the past. If a conditioned histogram for the transition $i\to j$ converges to a stationary distribution at some $k\geqslant n$, then the system only remembers up to $n$ steps in the past prior to the transition. The hidden microscopic paths projecting onto the states observed $k\geqslant n$ steps in the past thus mix in path space prior to the observed transition $i\to j$, which erases memory about the more distant past. 
The convergence rate of the histograms therefore informs about the temporal extent of \emph{local memory}.

We prove the convergence of the histograms under the assumption that there is at least one ``observable Markov state'' $\gamma$ in the network, while we do not need to know which state it is exactly. This is sufficient but not necessary for convergence.
There are systems that violate the assumption but still yield converging histograms (for an example see Appendix~\hyperref[sec:noMarkov]{H}). An ``observable Markov state'' $\gamma$ is an observed state corresponding to only one microscopic state (i.e., no hidden states corresponding to this observable state), satisfying the Markov property: for any integers $k\geqslant 1$ and $g\geqslant k+2$, for
any future sequence $\{s_{t}\}_{k+2\leqslant t\leqslant g}$ and history sequence $\{s_{t}\}_{1\leqslant t\leqslant k}$, 
\begin{align}
    \mp(\{s_{t}\}_{k+2\leqslant t\leqslant g}|s_{k+1}=\gamma;\{s_{t}\}_{1\leqslant t\leqslant k})  = \mp(\{s_{t}\}_{k+2\leqslant t\leqslant g}|s_{k+1}=\gamma) .
    \label{def_Markov_state}
\end{align}
We prove that the total variation distance (TVD) between the histogram $\mathbf{h}_{k}$ and the ``stationary'' histogram $\mathbf{h}_{\infty}$ is bounded by 
\begin{align}
    \Vert \mathbf{h}_{k} - \mathbf{h}_{\infty} \Vert_{\mathrm{TV}}
    = \frac{1}{2}\sum_{n=0}^L |h_k(n)-h_{\infty}(n)| 
    \leqslant C |\lambda^*|^{k+1-m^*}\binom{k+1}{m^*-1},
    \label{bound}   
\end{align}
where $n=p/2\delta$ is the index of bars in the histogram, $\lambda^*$
is the eigenvalue with the largest absolute value (excluding 1) for
a modified transition matrix $\mathbf{\Psi}$, $C$ is a constant
determined by the transition matrix, and $m^*$ is the dimension of the
largest Jordan block among all generalized eigenvalues of
$\mathbf{\Psi}$. The right hand side of Eq.~\eqref{bound} vanishes as
$k$ increases to infinity. Specifically, if $\mathbf{\Psi}$ is
  diagonalizable, then $m^*=1$ and the bound decays exponentially with
  power $k$ (see definition of $\mathbf{\Psi}$ in Eq.~\eqref{SM_psi_1}
  in Appendix \hyperref[sec:matrix]{A} and a detailed proof of the bound in Appendix \hyperref[sec:convg]{B}). Note that the bound does not depend on the choice of the bin width $\delta$.
 
Intuitively, if a history sequence includes the visit of an  observable Markov state, then the transition probability is fully  determined by the sequence after the visit and does \emph{not}  depend on the history further in the past. Therefore, as $k$ increases, the changing component of the histogram is due to  history sequences which do not visit a Markov state, the total  probability of which tends to zero. Specifically, if the observed sequence is an $n$-th (definite) order Markov process, i.e., if the future events are fully determined by $n$ steps in the past, then there exists an $n$ such that, for any $k\geqslant n$ and any $S^j_k$, $\mp(i|jS^j_k)=\mp(i|jS^j_n)$. In this case the histograms fully converge for $k\geqslant n$.

We now briefly discuss the concept of an $n$-th order Markov process.
Convergence of histograms suggests that we can also consider a ``weaker'' definition of an $n$-th order Markov process, where all the histograms for $S_n$ approach stationary distributions, i.e.\ changes become sufficiently small for $k>n$. In most realistic scenarios one encounters weak $n$-th order Markov processes. 

Notably, the concept of ``Markovian events''---events whose detection implies conditional independence between past and future time evolution---was proposed in \cite{Meer2023,Deguenther2024,Deguenther2024a}. Specifically, the visit of an observable Markov state is a Markovian event.
The assumption that such events can be identified turns out to provide a crucial improvement in thermodynamic inference. However, so far Eq.~\eqref{def_Markov_state} could not be generally tested in practice, as it is impossible to test for arbitrarily large $g$ and $k$ with a finite trajectory. Note that the Markov property at a specific state for a one-step transition
$\mp(s_{g+1}|s_{g}=\gamma;\{s_{t}\}_{1\leqslant t\leqslant g-1}) = \mp(s_{g+1}|s_{g}=\gamma)$ is only a necessary but \emph{not} sufficient condition for the Markov property. 
Here we provide a new tool to further test for a Markov state, though it is still not a complete solution. In the case of observable dynamics with a \emph{definite} order $n$, it is sufficient to test the Markovianity for no more than $n$ steps in the past and in the future (see Appendix \hyperref[sec:detection]{D} for details), which is practically feasible. Moreover, one can test for more than one first step in the future to get a more reliable (though still not definite) detection of a Markov state. Considering such extended future sequences also provides more information on the microscopic network (see Appendix~\hyperref[sec:cond_prob_two_step]{C}).

\section{Example}
Consider a toy model of a protein represented by the Markov network \cite{Prinz2011,bowman2013introduction,Olsson2017} in Figure~\ref{fig1}(a).
The microscopic states represent different stages of the folding of an idealized protein with 3 domains The $0/1$ represents the unfolding/folding of each domain. If it is not possible to resolve each microscopic state, but only to observe the number of folded domains, the $8$ states are projected onto $4$ observable states. The observable Markov states $a$ and $d$ represent the fully unfolded and folded state of the protein, respectively.
The transition rates $L_{\alpha\beta}$ for the Markov jump process are drawn randomly. In this example, the effective transition matrix $\mathbf{\Omega}$ is the same as $\mathbf{\Phi}$. The eigenvalues of the matrices are listed in Table~\ref{tab:table1}.
\begin{table}[h]
    \caption{Eigenvalues of the splitting probability matrix
      $\mathbf{\Phi}$ and the transition matrix $\mathbf{\Psi}$ with
      absorbing boundaries at states $a$ and $d$ (see the
        definition of $\mathbf{\Psi}$ in Eq.~\eqref{SM_psi_1} in Appendix \hyperref[sec:matrix]{A}). Values are rounded to 3 decimal places.}
    \label{tab:table1}
    \centering
    \begin{tabular}{ccccccccc}
    \hline
    & $\lambda_0$ & $\lambda_1$ & $\lambda_2$ & $\lambda_3$ & $\lambda_4$ & $\lambda_5$ & $\lambda_6$ & $\lambda_7$ \\
    \hline
    $\mathbf{\Phi}$ & $1$ & $-1$ & $0.792$ & $-0.792$ & $0.110$ & $-0.110$ & $0.053$ & $-0.053$ \\
    $\mathbf{\Psi}$ & $1$ & $1$ & $0.631$ & $-0.631$ & $0.528$ & $-0.528$ & $0.053$ & $-0.053$ \\
    \hline
    \end{tabular}
    
\end{table}

We simulate the microscopic Markov process to obtain one trajectory with $10^8$ transitions, which is sufficient for the statistics to sample histograms up to $k=12$ (see~Appendix~\hyperref[sec:sim]{E}~for simulation details). We project the full trajectory onto the observable space (states $a$, $b$, $c$, and $d$), and analyze the sequence of visited states. Note that, while our method can be readily applied to driven systems (see Appendix \hyperref[sec:driven]{F}) as long as the corresponding networks are irreducible with time-homogeneous transition rates, the example here satisfies detailed balance. 

\begin{figure}[h]
    \includegraphics[width=1.0\textwidth]{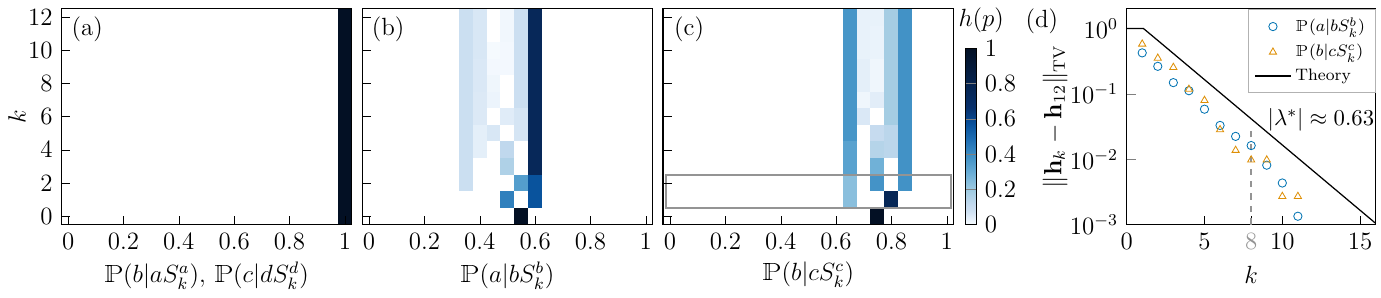}
    \caption{(a)-(c)~Histograms of transition probabilities between different pairs of states conditioned on history of different lengths $k$. Each row represents a histogram for a fixed history length $k$. The two rows in gray rectangle in panel (c) correspond to the histograms in Figure~\ref{fig1}(c-d).~The width of bars  in all histogram is $2\delta=0.05$. The color bar is the same for all plots. (d) The points show the TVD determined from simulation results, and the solid line is the theoretical bound in Eq.~\eqref{bound} which is independent of the pair of states. 
    }
    \label{fig2}
\end{figure}

Histograms of transition probabilities $\mp(i|jS^j_k)$ between different pairs of states conditioned on histories of different lengths $k$ are shown in Figure~\ref{fig2}(a-c). Specifically, the conditional probability $\mp(b|aS^a_k)$ and $\mp(c|dS^d_k)$ in Figure~\ref{fig2}(a) takes the value $1$ independent of the histories, because states $a$ and $d$ are both reflecting and Markov, i.e. the next step after $a$ is always $b$, and after $d$ always $c$. In this network, the histograms for $\mp(c|bS^b_k)$ (and $\mp(d|cS^c_k)$), which are not displayed, contain exactly the same information as those for $\mp(a|bS^b_k)$ ($\mp(b|cS^c_k)$) because $\mp(a|bS^b_k)=1-\mp(c|bS^b_k)$ holds for any $S^b_k$.

We take the transition probabilities $\mp(b|cS^c_k)$ in
Figure~\ref{fig2}(c) as an example of how to analyze the dynamics of
observed sequences in detail. The conditional probability irrespective
of history is one value $\mp(b|c)\approx 0.763$, determined by the number of occurrences of sequence $c\rightarrow b$ divided by that of state $c$, corresponding to the row at $k=0$. 
Adding one step in history, the histogram of $\mp(b|cS^c_1)$ exhibits two bars in Figure~\ref{fig1}(c), corresponding to the row $k=1$ in Figure~\ref{fig2}(c). The locations of the two bars are determined by the number of occurrences of sequence ${d\rightarrow c\rightarrow b}$ divided by that of ${d\rightarrow c}$, and $b\rightarrow c\rightarrow b$ by ${b\rightarrow c}$, respectively. Two distinct bars indicate that state $c$ is non-Markov, since the transition probability depends on the history. Therefore, state $c$ contains more than one microscopic state, and the transition ${c\rightarrow b}$ consists of at least two microscopic transition paths with different splitting probabilities.

We now increase the length of the history to $2$ steps as shown in Figure~\ref{fig1}(d). The bar due to $\mp(b|cdc)$ remains the same as in Figure~\ref{fig1}(c), because $d$ is an observable Markov state, i.e. $\mp(b|cdc) = \mp(b|cd)$. In contrast, the bar arising from $\mp(b|cb)$ in Figure~\ref{fig1}(c) splits into two in Figure~\ref{fig1}(d) because $b$ is not a Markov state.
As the history sequence $S^j_k$ becomes longer, the histograms include more information about the history, and more bars at different positions appear. When the history length $k$ becomes sufficiently large, the histogram converges to a stationary distribution. The dynamics of the observed sequence can be considered, \emph{locally} at state $c$, an $8$-th order Markov process in the weak sense of convergence with a cutoff $\Vert\mathbf{h}_{k} - \mathbf{h}_{\infty}\Vert < 10^{-2}$ for any $k\geqslant 8$. 

Furthermore, in the example in Figure~\ref{fig2}(c) for transition  $c\rightarrow b$ we can infer, from the location of the bars, that there is at least one element of the splitting probability from a microscopic state $\aa$ in $c$ to the observable state $b$, $\mp(b|\aa) \equiv \sum_{\beta}\! \ind{b}[\beta] \mp(\beta|\aa)$, smaller than $0.65$ and another one larger than $0.85$. The true minimum and maximum values of splitting probabilities are bounded by $\mp(b|7)\approx 0.59<0.65$ and $\mp(b|6)\approx0.91>0.85$.

A comparison between the TVD from simulation and the theoretical bound is shown in Figure~\ref{fig2}(d). The TVD from simulation is determined by $\Vert \mathbf{h}_{k} - \mathbf{h}_{k_{\mathrm{max}}} \Vert_{\mathrm{TV}} \leqslant 2\Vert \mathbf{h}_{k} - \mathbf{h}_{\infty} \Vert_{\mathrm{TV}}$. In agreement with Eq.~\eqref{bound}, the histograms converge to stationary histograms for large $k$. 

\section{Discussion}
A first application of the method is to test the consistency of
Markov-state models. It is common to model single-molecule
trajectories with a Markov transition network
\cite{Pande,Chodera}. For fitted trajectories, we can build histograms
for the transition probabilities conditioned on different lengths of
history. If a Markov model indeed captures all relevant states,
i.e.~does not leave any hidden paths between observed states, then
histograms between any pair of states should display a single bar
which remains the same for any history length. If the trajectories are
not long enough, or their number is not large enough, the probability
distributions in the histograms may evolve (i.e.~spread) as the
history length increases due to undersampling, while the distribution
should concentrate around the same
value~(see~Appendix~\hyperref[sec:sampling]{G}~for a discussion of
artifacts due to insufficient statistics). However, if the histograms
for a specific pair of states exhibit more than one peak that changes
with the history length, this \emph{conclusively} implies hidden paths
between said pair of states that are \emph{not} captured in the Markov
model. Moreover, if we were to choose a very small bin width for  the histogram, we would also observe a finer-grained distribution of the transition probabilities due to finite statistics. The discussion in Appendix~\hyperref[sec:sampling]{G} applies also in this case. Nevertheless, for systems with observable Markov states, the histograms converge regardless of the choice of bin width because the convergence is guaranteed by the theoretical bound in Eq.~\eqref{bound} which does not depend on $\delta$.

Current experimental trajectories \cite{Noe2008,Stigler2011,Vollmar2024} are sufficiently long to detect the emergence of memory \emph{locally}, which typically unravels already with a small $k\geqslant 1$.
If, however, an observed trajectory  is not long enough to determine the exact duration of memory, i.e.,~to see the convergence,  we can construct histograms up to some maximal $k_{\mathrm{max}}$ given the statistics, i.e.,\ every possible distinct history should be sampled at least a few times. We can assess the convergence qualitatively by comparing histograms as in Figure~\ref{fig2}(a-c), or quantitatively from the decay of TVD as in Figure~\ref{fig2}(d). If the histograms do not converge, the duration of local memory is longer than $k_{\mathrm{max}}$. The non-convergence is either because there is \emph{no} observable Markov state in the network, or because the transition is so far from a Markov state that it takes more than $k_{\mathrm{max}}$ steps of the preceding history sequence to visit the Markov state with a significant probability.

Moreover, since observed transition probabilities are always linear combinations of effective microscopic splitting probabilities $\mp(i|\aa_{k+1})$, as shown in Eq.~\eqref{def_split_prob1}, 
the locations of bars are always within the range of the minimum and maximum of the effective microscopic splitting probabilities. Based on this property, the histograms also provide information on the range of effective hidden splitting probabilities.  
Therefore, the locations of bars in histograms provide information on hidden splitting probabilities between a given pair of states, and the convergence of the respective histograms reveals the extent of memory \emph{locally}.

A continuous-time jump process in our setting is fully described by the transition probability between pairs of mesostates, i.e. observable states, and the waiting time in the mesostate until the transition into another given mesostate, both conditioned on the history of visited states. The waiting time in the current mesostate until the transition into another given mesostate depends only on the mesostates in previous steps but \emph{not} the waiting time in the previous mesostates. Therefore, our method may be extended to account for waiting times in the respective mesostates by looking at the waiting time distribution conditioned on history sequences of different length.

Our method may be applicable to trajectories  from single-molecule experiments  \cite{Pirchi2011,Scheiderer2025}, which, due to an inherent projection and a finite temporal resolution,  typically do not distinguish all microscopic conformations, and to chemically reacting systems \cite{Gillespie2007,Frezzato2025} where  only the total copy-number of molecules is observed. The proposed method may  be complementary to existing methodology \cite{Kolomeisky2007,Moffitt2014,Vollmar2024,Berezhkovskii2018,Lapolla2021,Song2023,Meer2023}.
Our work might also aid the inference on microscopic  dynamics using other techniques, such as Hidden Markov Models and Machine Learning \cite{Pirchi2011,Noe2013,Mardt2018,Holmstrom2019,Mardt2022,Tancredi2024}.

\section{Conclusion}
We proposed a histogram-based method to infer, for the first time, memory \emph{locally}, i.e.~non-Markovianity of transitions between any pair of observed states. The method is based on transition probabilities conditioned on history sequences of different lengths. The convergence rate of these ``history'' histograms towards a stationary distribution provides, for the first time, \emph{a local quantification of the duration of memory}, which reflects how distinct microscopic paths projecting onto the same observed transition decorrelate in path space. The duration of memory is measured by the number of observable transitions $k$ prior to a specific transition. Accordingly, this provides information about the hidden topology of microscopic paths in a holography-like fashion. 
Under the assumption of at least one observable Markov state, we proved that the histograms indeed always converge to a stationary distribution when the history becomes infinitely long. 
The analytical predictions were corroborated by simulation results for a $8$-state Markov network example.   

The histogram-based method provides information on local (hidden) properties of the network, which is quite different from (but complementary to) the information gained from local waiting time distributions. 
The method can be applied to test whether a Markov-state model consistently captures the underlying dynamics and also provides a more reliable (yet still \emph{not} definitive) practical criterion for a ``Markov state''. It would be interesting to connect the local memory inferred here to properties of memory kernels \cite{Tanja_PR,Tanja_JCP,Tanja_3,Ginot2022,Roland_PRL,Lavacchi_2024,Dalton_2025}
and to seek for a holography-like understanding of the weak-memory regime \cite{Kay_1,Kay_2}.

\section*{Acknowledgments}
Financial support from the Volkswagen Stiftung together with the Ministry of Science and Culture of the German State of Lower Saxony within the scope of the ``Research Cooperation Lower Saxony - Israel'' (to A.G.), the European Research Council (ERC) under the European Union’s Horizon Europe research and innovation program (Grant Agreement No. 101086182 to A.G.), the US National Science Foundation (Grant no. CHE 2400424 to D.E.M.),  the Robert A. Welch Foundation (Grant no. F-1514 to D.E.M.) and the Alexander von Humboldt Foundation is gratefully acknowledged.

\section*{Data availability statement}
The data that support the findings of this study are openly available at the following URL: https://doi.org/10.5281/zenodo.15853642.

\newpage

\begin{appdxsection}
 
\section*{Appendix A:~Effective transition matrix} \label{sec:matrix}
\renewcommand{\theequation}{A.\arabic{equation}}
\setcounter{equation}{0}

In this appendix we introduce an effective transition matrix $\mathbf{\Omega}$ for the reduced sequence $\{\aa_{t}\}_{1\leqslant t\leqslant k+2}$ consisting of the first microscopic steps after each observable transition. In the main text, we defined the observed sequence $\{s_{t}\}_{1\leqslant t\leqslant k+2}$ and a corresponding microscopic (fully resolved) sequence $\{\sigma_{\tau}\}_{1\leqslant \tau\leqslant k'}\,(k'\geqslant k+2)$.

Assume that the system has $m$ microscopic states $\{1,\ldots,m\}$ which are projected onto $n$ observable states $\{1,\ldots,n\}$. The microscopic states are enumerated in an order such that those projected onto the same observable state are adjacent. We denote the number of microscopic states in an observable state $a$ as $r_a$, satisfying $\sum_{a=1}^{n} r_a=m$.


As an example we first consider a one-step transition. Assume the current microscopic state $\aa_{t}$ is projected to an observable state $s_t=s$. We define for short that $g \equiv \sum_{a=1}^{s-1} r_a$ and $r\equiv r_s$, i.e. the observable state $s$ consists of $r$ microscopic states.
We partition the transition probability matrix $\mathbf{\Phi}$ into block matrices by separating the columns corresponding to the observable state $s$, as
\begin{align}
    \mathbf{\Phi} = \left(
    \begin{array}{c  c  c}
        \cline{3-3}
        \mathbf{\Phi}_{11}^{g\times g} & \mathbf{R}^{g\times r} & \multicolumn{1}{|c|}{} \\
        \cline{1-1}
        \multicolumn{1}{|c|}{} & \mathbf{Q}^{r\times r} & \multicolumn{1}{|c|}{\raisebox{.5\normalbaselineskip}[0pt][0pt]{$\mathbf{\Phi}_{13}^{(g+r)\times (m-g-r)}$}} \\
        \cline{3-3}
        \multicolumn{1}{|c|}{\raisebox{.5\normalbaselineskip}[0pt][0pt]{$\mathbf{\Phi}_{31}^{(m-g)\times g}$}} & \mathbf{R}'^{(m-g-r)\times r} & \mathbf{\Phi}_{33}^{(m-g-r)\times(m-g-r)} \\
        \cline{1-1}
    \end{array}
    \right),
\end{align}
where the superscripts indicate the dimensions of the respective block matrices.
Since we do \emph{not} distinguish the microscopic full sequences which stay in the observable state $s$ for arbitrary steps before jumping to another observable state, we
consider the sum of the probabilities of all microscopic sequences
$\{\sigma_{\tau}\}_{\tau(t)+1\leqslant
  \tau'\leqslant\tau(t+1)-1}$ which stay in the observable
state $s_{t}$ until the next observable step $t+1$. Therefore, the
transition probability
$\Omega_{\aa_{t+1}\aa_{t}}=\mp(\aa_{t+1}|\aa_{t})$ for the reduced
sequence given by 
\begin{align}
    &\mp(\aa_{t+1}|\aa_{t}) \nonumber\\ =& \sum_{\tau(t+1)=\tau(t)+1}^{\infty} \sum_{ \{\sigma_{\tau}\}_{\tau(t)+1\leqslant \tau'\leqslant\tau(t+1)-1} } \!\! \mp(\sigma_{\tau(t+1)}=\aa_{t+1};\,\{\sigma_{\tau'}\}_{\tau(t)+1\leqslant \tau'\leqslant\tau(t+1)-1} | \sigma_{\tau(t)}=\aa_{t}) \nonumber\\
    =& \sum_{\tau(t+1)=\tau(t)+1}^{\infty} \sum_{\sigma_{\tau(t)+1}=g+1}^{g+r} \sum_{\sigma_{\tau(t)+2}=g+1}^{g+r} \cdots \sum_{\sigma_{\tau(t+1)-1}=g+1}^{g+r} \underbrace{\Phi_{\aa_{t+1}\sigma_{\tau(t+1)-1}}}_{\scalebox{0.7}{$\quad\;\,\displaystyle\begin{pmatrix}\mathbf{R}\\\mathbf{0}\\\mathbf{R}'\end{pmatrix}$}} \underbrace{\prod_{\tau'=\tau(t)}^{\tau(t+1)-2} \Phi_{\sigma_{\tau'+1}\sigma_{\tau'}}}_{\mathbf{Q}^{\tau(t+1)-\tau(t)-1}} .
    \label{SM_sum_seq_prob}
\end{align}
Note that when $\tau(t+1)=\tau(t)+1$, there is no summation or
product, i.e.\ the first term is given by
$\Phi_{\aa_{t+1}\aa_{t}}\mp(\aa_{t})$. 
For example, if we consider $t=1$, then $\tau(1)=1$, and Eq.~\eqref{SM_sum_seq_prob} becomes
\begin{align}
    \Phi_{\aa_{2}\aa_{1}} \mp(\aa_{1}) + \sum_{\tau(2)=3}^{\infty} \sum_{\sigma_{2}=g+1}^{g+r} \cdots \sum_{\sigma_{\tau(2)-1}=g+1}^{g+r} \Phi_{\aa_{2}\sigma_{\tau(2)-1}} \prod_{\tau'=2}^{\tau(2)-2} \Phi_{\sigma_{\tau'+1}\sigma_{\tau'}} \Phi_{\sigma_{2}\aa_{1}} \mp(\aa_{1}).
\end{align}
With a change of variable $d\equiv \tau(t+1)-\tau(t)-1$, Eq.~\eqref{SM_sum_seq_prob} can be written in the matrix formalism. For any $\aa_{t+1}\in\{1,\dots,g\}\cup\{g+r+1,\dots,m\}$ and $\aa_{t}\in\{g+1,\dots,g+r\}$,
\begin{align}
    \Omega_{\aa_{t+1}\aa_{t}} = \mp(\aa_{t+1}|\aa_{t}) = \left( 
    \begin{pmatrix}
        \mathbf{R} \\ \mathbf{0} \\ \mathbf{R}'
    \end{pmatrix}
    \sum_{d=0}^{\infty} \mathbf{Q}^d \right)_{\aa_{t+1}\aa_{t}} = \left( 
    \begin{pmatrix}
        \mathbf{R} \\ \mathbf{0} \\ \mathbf{R}'
    \end{pmatrix} 
    (\mathbf{I}-\mathbf{Q})^{-1}
    \right)_{\aa_{t+1}\aa_{t}}.
    \label{SM_omega_1}
\end{align}
We used the identity $\sum_{d=0}^{\infty} \mathbf{Q}^d = (\mathbf{I}-\mathbf{Q})^{-1}$. Matrix $\mathbf{I}-\mathbf{Q}$ is always invertible because $\mathbf{Q}$ has a spectral radius less than $1$.
To determine all the elements in $\mathbf{\Omega}$, we introduce a matrix $\mathbf{\Lambda}_{s}$ by setting all the states other than the microscopic states in $s$ to absorbing states, yielding in a transition matrix
\begin{align}
    \mathbf{\Lambda}_{s} \equiv \left(
    \begin{array}{c  c  c}
        \cline{3-3}
        \mathbf{I}^{g\times g} & \mathbf{R}^{g\times r} & \multicolumn{1}{|c|}{} \\
        \cline{1-1}
        \multicolumn{1}{|c|}{} & \mathbf{Q}^{r\times r} & \multicolumn{1}{|c|}{\raisebox{.5\normalbaselineskip}[0pt][0pt]{$\mathbf{0}^{(g+r)\times (m-g-r)}$}} \\
        \cline{3-3}
        \multicolumn{1}{|c|}{\raisebox{.5\normalbaselineskip}[0pt][0pt]{$\mathbf{0}^{(m-g)\times g}$}} & \mathbf{R}'^{(m-g-r)\times r} & \mathbf{I}^{(m-g-r)\times(m-g-r)} \\
        \cline{1-1}
    \end{array}
    \right).
    \label{SM_lambda_1}
\end{align}
It is easy to show that the $d$-th power of the matrix $\mathbf{\Lambda}_{s}$ is given by
\begin{align}
    \mathbf{\Lambda}_{s}^{d} =  \left(
    \begin{array}{c  c  c}
        \cline{3-3}
        \mathbf{I}^{g\times g} & \mathbf{R}(\mathbf{I}^{r\times r}-\mathbf{Q})^{-1}(\mathbf{I}^{r\times r}-\mathbf{Q}^{d}) & \multicolumn{1}{|c|}{} \\
        \cline{1-1}
        \multicolumn{1}{|c|}{} & \mathbf{Q}^{d} & \multicolumn{1}{|c|}{\raisebox{.5\normalbaselineskip}[0pt][0pt]{$\mathbf{0}^{(g+r)\times (m-g-r)}$}} \\
        \cline{3-3}
        \multicolumn{1}{|c|}{\raisebox{.5\normalbaselineskip}[0pt][0pt]{$\mathbf{0}^{(m-g)\times g}$}} & \mathbf{R}'(\mathbf{I}^{r\times r}-\mathbf{Q})^{-1}(\mathbf{I}^{r\times r}-\mathbf{Q}^{d}) & \mathbf{I}^{(m-g-r)\times(m-g-r)} \\
        \cline{1-1}
    \end{array}
    \right).
\end{align}
Therefore, the infinite power of $\mathbf{\Lambda}_{s}$ is given by
\begin{align}
    \mathbf{\Lambda}_{s}^{\infty}\equiv \lim_{d\to\infty}\mathbf{\Lambda}_{s}^{d} =  \left(
    \begin{array}{c  c  c}
        \cline{3-3}
        \mathbf{I}^{g\times g} & \mathbf{R}(\mathbf{I}^{r\times r}-\mathbf{Q})^{-1} & \multicolumn{1}{|c|}{} \\
        \cline{1-1}
        \multicolumn{1}{|c|}{} & \mathbf{0}^{r\times r} & \multicolumn{1}{|c|}{\raisebox{.5\normalbaselineskip}[0pt][0pt]{$\mathbf{0}^{(g+r)\times (m-g-r)}$}} \\
        \cline{3-3}
        \multicolumn{1}{|c|}{\raisebox{.5\normalbaselineskip}[0pt][0pt]{$\mathbf{0}^{(m-g)\times g}$}} & \mathbf{R}'(\mathbf{I}^{r\times r}-\mathbf{Q})^{-1} & \mathbf{I}^{(m-g-r)\times(m-g-r)} \\
        \cline{1-1}
    \end{array}
    \right),
\end{align}
which contains the transition probabilities from the microscopic states in the observable state $s$ to other microscopic states, i.e.\ Eq.~\eqref{SM_omega_1} can be expressed from from $\mathbf{\Lambda}_{s}^{\infty}$.
 
More formally, we define the matrix to select the first and the third column (row) of a $3\times 3$ block matrix as
\begin{align}
    \mathbf{U}_{s}^{(m-r)\times m} \equiv 
    \begin{pmatrix}
        \mathbf{I}^{g\times g} & \mathbf{0}^{g\times r} & \mathbf{0}^{g\times(m-g-r)} \\
        \mathbf{0}^{(m-g-r)\times g} & \mathbf{0}^{(m-g-r)\times r} & \mathbf{I}^{(m-g-r)\times(m-g-r)} \\
    \end{pmatrix},
\end{align}
and the matrix to select the second column (row) as
\begin{align}
    \mathbf{V}_{s}^{r\times m} \equiv 
    \begin{pmatrix}
        \mathbf{0}^{r\times g} & \mathbf{I}^{r\times r} & \mathbf{0}^{r\times(m-g-r)} \\
    \end{pmatrix}.
\end{align}
Then we can formally express $\mathbf{\Lambda}_{s}$ defined in Eq.~\eqref{SM_lambda_1} as
\begin{align}
    \mathbf{\Lambda}_{s} = \mathbf{U}_{s}^\top\mathbf{U}_{s} + \mathbf{U}_{s}^\top\mathbf{R}_{s}\mathbf{V}_{s} + \mathbf{V}_{s}^\top\mathbf{Q}_{s}\mathbf{V}_{s},
\end{align}
where $\mathbf{R}_{s} = \mathbf{U}_{s}\mathbf{\Phi}\mathbf{V}_{s}^\top$, $\mathbf{Q}_{s} = \mathbf{V}_{s}\mathbf{\Phi}\mathbf{V}_{s}^\top$.
The sets of microscopic states projected onto different observable
states are 
disjoint. Thus, we can carry out the same calculation for all $s\in\{1,\ldots , n\}$ and define
\begin{align}
    \mathbf{\Lambda} \equiv \sum_{s=1}^{n} \left( \mathbf{U}_{s}^\top\mathbf{R}_{s}\mathbf{V}_{s} + \mathbf{V}_{s}^\top\mathbf{Q}_{s}\mathbf{V}_{s} \right).
\end{align}
Therefore, the new ``effective'' transition matrix is given by the infinite power of $\mathbf{\Lambda}$ as
\begin{align}
    \mathbf{\Omega}\equiv\mathbf{\Lambda}^{\infty} = \sum_{s=1}^{n} \mathbf{U}_{s}^\top \mathbf{R}_{s} (\mathbf{I}-\mathbf{Q}_{s})^{-1}\mathbf{V}_{s}.
    \label{SM_omega_2}
\end{align} 
Every column of $\mathbf{\Omega}$ adds up to $1$, and the diagonal
blocks are all-zero matrices, agreeing with the fact that there is no
``self-loop'' in the observable sequence. Every block matrix
corresponds to the microscopic splitting probabilities between two
observable states. We can write $\mathbf{\Omega}$ more explicitly in
block form as
\begin{align}
    \mathbf{\Omega} =  \left(
    \begin{array}{c c c c}
        \cline{4-4} 
        \mathbf{0}^{r_1\times r_1} & \mathbf{R}_{2}(\mathbf{I}^{r_2\times r_2}-\mathbf{Q}_2)^{-1} & \dots & \multicolumn{1}{|c|}{} \\
        \cline{1-1}
        \multicolumn{1}{|c|}{} & \mathbf{0}^{r_2\times r_2} & \dots & \multicolumn{1}{|c|}{\raisebox{.0\normalbaselineskip}[0pt][0pt]{$\mathbf{R}_{n}(\mathbf{I}^{r_n\times r_n}-\mathbf{Q}_n)^{-1}$}} \\
        \cline{2-2}
        \multicolumn{1}{|c|}{\raisebox{.0\normalbaselineskip}[0pt][0pt]{$\mathbf{R}_{1}(\mathbf{I}^{r_1\times r_1}-\mathbf{Q}_1)^{-1}$}} & \multicolumn{1}{|c|}{} & \ddots & \multicolumn{1}{|c|}{} \\
        \cline{4-4}
        \multicolumn{1}{|c|}{} & \multicolumn{1}{|c|}{\raisebox{.5\normalbaselineskip}[0pt][0pt]{$\mathbf{R}_{2}'(\mathbf{I}^{r_2\times r_2}-\mathbf{Q}_2)^{-1}$}} & \dots & \mathbf{0}^{r_{n}\times r_{n}} \\
        \cline{1-2}
    \end{array}
    \right).
\end{align}
We mentioned in the main text that the positions of non-zero bars in the histograms are bounded by the smallest and largest values of $\Omega_{\aa_{k+2}\aa_{k+1}}$. More precisely, the positions of non-zero bars in the histograms for $\mp(i|jS_k)$ are bounded by the smallest and largest elements in the block matrix 
\begin{align}
    \left(
    \begin{array}{c}
        \mathbf{R}_{j}(\mathbf{I}^{r_j\times r_j}-\mathbf{Q}_j)^{-1}  \\
        \mathbf{R}_{j}'(\mathbf{I}^{r_j\times r_j}-\mathbf{Q}_j)^{-1}
    \end{array}
    \right),
\end{align}
which consists of the outward transition probabilities from microscopic states in $j$.

For the proof of convergence in Appendix~\hyperref[sec:convg]{B}, we will assume the existence of $q\geqslant 1$ observable Markov state. For notational convenience, we enumerate microscopic states such that the first $q$ states are observable. Then we will define a transition matrix $\mathbf{\Psi}$ with the observable Markov states set as absorbing states, and the other elements the same as in $\mathbf{\Omega}$
\begin{align}
    \mathbf{\Psi} \equiv \sum_{s=1}^{q}\mathbf{e}_{s}\mathbf{e}_{s}^\top + \sum_{s=q+1}^{n} \mathbf{U}_{s}^\top \mathbf{R}_{s} (\mathbf{I}-\mathbf{Q}_{s})^{-1}\mathbf{V}_{s},
    \label{SM_psi_1}
\end{align}
where we define $\mathbf{e}_s$ as the column vector with the $s$-th element being $1$ and elsewhere $0$.
We can write $\mathbf{\Psi}$ more explicitly in block form as
\begin{align}
    \mathbf{\Psi} =  \left(
    \begin{array}{c c c c}
        \cline{4-4} %
        \mathbf{I}^{q\times q} & \mathbf{R}_{q+1}(\mathbf{I}^{r_{q+1}\times r_{q+1}}-\mathbf{Q}_{q+1})^{-1} & & \multicolumn{1}{|c|}{} \\
        \cline{1-1}
        \multicolumn{1}{|c|}{} & \mathbf{0}^{r_{q+1}\times r_{q+1}} & \multicolumn{1}{c}{\raisebox{.0\normalbaselineskip}[0pt][0pt]{$\ddots$}} & \multicolumn{1}{|c|}{} \\
        \cline{2-2}
        \multicolumn{1}{|c|}{} & \multicolumn{1}{|c|}{} & \multicolumn{1}{c}{\raisebox{-.5\normalbaselineskip}[0pt][0pt]{$\ddots$}} & \multicolumn{1}{|c|}{\raisebox{.5\normalbaselineskip}[0pt][0pt]{$\mathbf{R}_{n}(\mathbf{I}^{r_n\times r_n}-\mathbf{Q}_n)^{-1}$}} \\
        \multicolumn{1}{|c|}{\raisebox{.5\normalbaselineskip}[0pt][0pt]{$\mathbf{0}^{(m-1)\times q}$}} & \multicolumn{1}{|c|}{\raisebox{.0\normalbaselineskip}[0pt][0pt]{$\mathbf{R}_{q+1}'(\mathbf{I}^{r_{q+1}\times r_{q+1}}-\mathbf{Q}_{q+1})^{-1}$}} & & \multicolumn{1}{|c|}{} \\
        \cline{4-4}
        \multicolumn{1}{|c|}{} & \multicolumn{1}{|c|}{} & \multicolumn{1}{c}{\raisebox{.0\normalbaselineskip}[0pt][0pt]{$\ddots$}} & \mathbf{0}^{r_{n}\times r_{n}} \\
        \cline{1-2}
    \end{array}
    \right).
    \label{SM_psi_2}
\end{align}

    We assume the Markov chain $\mathbf{\Phi}$ is irreducible, i.e. every state communicates with every other state. Thus, after setting the observable Markov states to absorbing states, all the remaining states become transient because they are accessible to an absorbing state.
    The transition matrix $\mathbf{\Psi}$ with $q$ absorbing states defined in Eq.~\eqref{SM_psi_1} can be written as
    \begin{align}
        \mathbf{\Psi} = 
        \begin{pmatrix}
            \mathbf{I}^{q\times q} & \mathbf{R}^{q\times (m-q)} \\
            \mathbf{0}^{(m-q)\times q} & \mathbf{Q}^{(m-q)\times (m-q)}
        \end{pmatrix},
        \label{SM_psi_3}
    \end{align}
    where $\mathbf{R}^{q\times (m-q)} \neq \mathbf{0}^{q\times (m-q)}$ because no states are isolated.

\section*{Appendix B:~Convergence of histograms} \label{sec:convg}
\renewcommand{\theequation}{B.\arabic{equation}}
\setcounter{equation}{0}
    In this appendix, we prove that the histograms of transition probabilities conditioned on histories converge to a stationary distribution when the length of histories tends to infinity. 
    
    As introduced in the main text, we assume that there is at least one observable Markov state in the network and denote the set of observable Markov states as $O=\{1,\ldots,q\}$. We split all possible history sequences $S_k$ of length $k$ into two disjoint sets depending on whether they have visited any observable Markov
    state or not, i.e.~$\{S^j_k\} = \{S^j_k\}^\mathrm{M}
    \cup \{S^j_k\}^\mathrm{N}$, where  
         \begin{align}
        \{S^j_k\}^\mathrm{M} &\equiv \{S^j_k\,|\, \exists\, t,\, 1\leqslant t\leqslant k :\, s_{t}\in O\}; \\
        \{S^j_k\}^\mathrm{N} &\equiv \{S^j_k\,|\, \forall\, t,\, 1\leqslant t\leqslant k:\, s_{t}\not\in O\}.
         \end{align}       
    The histogram is also decomposed into two parts $h_k(p) = h_k^\mathrm{M}(p) + h_k^\mathrm{N}(p)$. We will investigate the two parts separately in the following subsections. 
 
\subsection{Contribution from trajectories that visited the Markov
  state}\label{sec:convg_SkM} 
    In this subsection, we determine the contribution of $S^j_k\in\{S^j_k\}^\mathrm{M}$ to the histogram $h(p)$ from $\mp(i|jS^j_k)\in \left[p-\delta,p+\delta\right)$. The height of the bar contributed from $\{S^j_k\}^\mathrm{M}$ at $p$ is given by
    \begin{align}
        h_k^\mathrm{M}(p) = \frac{1}{\mp(j)} \sum_{S^j_k\in \{S^j_k\}^\mathrm{M}} \mp(jS^j_k)\mathbbm{1}_{[p-\delta,p+\delta)}[\mp(i|jS^j_k)].
    \end{align} 
    For any $S^j_k\in\{S^j_k\}^\mathrm{M}$, there exists the latest time step $t_m$ where $s_{t_m}=\aa_{t_m}\in O,\,s_{t>t_m}\not\in O$. Due to the definition of an observable Markov state, we can decompose the observable sequence probability
    $\mp(\aa_{k+1}S^j_k)$ into   
    \begin{align}
    	\mp(\aa_{k+1}S^j_k) = \mp(\aa_{k+1}\{s_{t}\}_{t_m+1\leqslant t\leqslant k}|s_{t_m}\in O) \mp(\{s_{t}\}_{1\leqslant t\leqslant t_m}).
     \label{SM_seq_prob_dec}
    \end{align}
   We can substitute Eq.~\eqref{SM_seq_prob_dec} into the definition of the conditional probability in Eq.~(2) of the main text to find
     \begin{align}
        \mp(i|jS^j_k) 
        &= \frac{\sum_{\aa_{k+1}}\!\! \ind{j}[\aa_{k+1}] \mp(\aa_{k+1}S^j_k) \mp(i|\aa_{k+1})}{\sum_{\aa_{k+1}} \ind{j}[\aa_{k+1}] \mp(\aa_{k+1}S^j_k)} \\
        &= \frac{\sum_{\alpha_{k+1}}\!\! \ind{j}[\alpha_{k+1}] \mp(\aa_{k+1}\{s_{t}\}_{t_m+1\leqslant t\leqslant k}|s_{t_m}\in O) \mp(i|\alpha_{k+1})}{\sum_{\alpha_{k+1}}\!\! \ind{j}[\alpha_{k+1}] \mp(\aa_{k+1}\{s_{t}\}_{t_m+1\leqslant t\leqslant k}|s_{t_m}\in O)} \\
        &= \mp(i|j\{s_{t}\}_{t_m+1\leqslant t\leqslant k}\gamma),
        \label{SM_split_prob}
     \end{align}
    where ${\mp(i|\aa_{k+1}) = \sum_{\aa_{k+2}}\!\! \ind{i}[\aa_{k+2}] \mp(\aa_{k+2}|\aa_{k+1})}$. Eq.~\eqref{SM_split_prob} is an important step in our calculation: the probability of early histories $\mp(\{s_{t}\}_{1\leqslant t\leqslant t_m})$ is cancelled because it does not depend on $k$ or $\aa_{k+1}$, i.e.~it is the same in every term in the summation. Therefore, the $\mp(i|jS^j_k)$ is determined only by the steps after $t_m$ in the history, i.e.~by $\{s_{t}\}_{t_m\leqslant t\leqslant k+2}$. Intuitively, this is because the memory resets after visiting the observable Markov state.
    
    We can extend the history $l$ steps further to the past to be $S^j_{k+l}\equiv\{s_{t}\}_{-l+1\leqslant t\leqslant k},\, l\geqslant 1$,
    and define $\{S^j_{k+l}\}^\mathrm{M}$ as the trajectories that do
    \emph{not} visit $\gamma$ in the last $k$ steps during $\{s_{t}\}_{1\leqslant t\leqslant k}$. The contributions to the histogram from $\{S^j_k\}^\mathrm{M}$ and $\{S^j_{k+l}\}^\mathrm{M}$ are exactly the same, for they have the same possible sequences $\{s_{t}\}_{1\leqslant t\leqslant k+2}$. Thus, for any $S^j_{k+l}=\{s_{t}\}_{-l+1\leqslant t\leqslant k} \in \{S^j_{k+l}\}^\mathrm{M}$, we have $\mp(i|j\{s_{t}\}_{-l+1\leqslant t\leqslant k}) = \mp(i|j\{s_{t}\}_{1\leqslant t\leqslant k})$.
    Since $\mp(i|jS^j_k)$ is only affected by the last $(k-t_m)$ steps of the history, $h_k^\mathrm{M}(p)$ also does not depend on the earlier history. We can determine the height of bars for $\{S^j_{k+l}\}^\mathrm{M}$ as
         \begin{align}
        h_{k+l}^\mathrm{M}(p) &= \frac{1}{\mp(j)} \sum_{\{s_{t}\}_{1\leqslant t\leqslant k}\in \{S^j_k\}^\mathrm{M}} \sum_{\{s_{t}\}_{-l+1\leqslant t\leqslant 0}}\!\! \mp(jS^j_{k+l}) \mathbbm{1}_{[p-\delta,p+\delta)}[\mp(i|jS^j_{k+l})] \\
        &= \frac{1}{\mp(j)} \sum_{\{s_{t}\}_{1\leqslant t\leqslant k}\in \{S^j_k\}^\mathrm{M}}\!\! \mp(jS^j_k) \mathbbm{1}_{[p-\delta,p+\delta)}[\mp(i|jS^j_k)] 
        = h_k^\mathrm{M}(p).
         \end{align}
    We have proved that the contribution from sequences $S^j_k\in\{S^j_k\}^\mathrm{M}$ is already stationary, i.e.\ does not change in the histograms for longer histories.

\subsection{Contribution from trajectories that did not visit the Markov state}\label{sec:convg_SkN}
    In this subsection, we prove that the total contribution from $S^j_k\in\{S^j_k\}^\mathrm{N}$ to the histogram with $L$ bars
    \begin{align}
        \sum_{z=0}^L h_k^\mathrm{N}(p=2z\delta) = \frac{1}{\mp(j)} \sum_{S^j_k\in\{S^j_k\}^\mathrm{N}}\!\! \mp(jS^j_k)
    \end{align}
    vanishes as $k$ tends to infinity, where we expressed the positions of bars as $p=2z\delta$, $0\leqslant z \leqslant L$, $2L\delta=1$.

    Trajectories with $\aa_{k+1}\in O=\{1,\ldots,q\}$ are trivial because the history $S^j_k$ does not matter. Thus, in the following calculation, we choose $j$ \emph{not} being an observable Markov state, i.e. $\aa_{k+1}\not\in O$. 
    For notational convenience (and w.l.o.g.), we enumerate microscopic states such that the observable Markov state $\gamma=1$. We introduce the transition matrix $\mathbf{\Psi}$ with the observable Markov states set as absorbing states, as in Eqs.~(\ref{SM_psi_1},\ref{SM_psi_2}).
    
     We define column vectors $\mathbf{e}_i \equiv (0\,\ldots\,0\,1\,0\, \ldots\,0)^\top$ with the $i$-th element being $1$, and the steady state distribution of $\mathbf{\Omega}$ as $\bm{\pi}$, which satisfies $\bm{\Omega}\bm{\pi} = \bm{\pi}$.
    The probability of starting from a steady state distribution $\bm{\pi}$, ending in $\aa_{k+1}$, and not visiting any observable Markov state between step $1$ and $k+1$ is given by 
    \begin{align}
        \mp(\aa_{k+1};\forall\, t\in[1,k+1],\, \aa_{t}\not\in O)=\mathbf{e}_{\aa_{k+1}}^\top\mathbf{\Psi}^{k+1}\bm{\pi}.
    \end{align}
    Therefore, we have
    \begin{align}
        \frac{1}{\mp(j)} \sum_{S^j_k\in\{S^j_k\}^\mathrm{N}}\!\! \mp(jS^j_k) &= \frac{1}{\pi(j)}\sum_{\aa_{k+1}}\ind{j}[\aa_{k+1}] \mp(\aa_{k+1};\forall\, t,\, \aa_{t}\not\in O) \nonumber\\
        &= \frac{1}{\pi(j)} \sum_{\aa_{k+1}}\ind{j}[\aa_{k+1}] \mathbf{e}_{\aa_{k+1}}^\top \mathbf{\Psi}^{k+1}\bm{\pi}.
        \label{SM_prob_sum}
    \end{align}
    Since $\mathbf{\Psi}^{k+1}\bm{\pi}$ has non-negative components and $\aa_{k+1}\notin O$, we can bound the projecting sum with a sum over all the states which are not observable Markov states, and obtain an inequality 
    
    \begin{align}
        \sum_{\aa_{k+1}}\ind{j}[\aa_{k+1}] \mathbf{e}_{\aa_{k+1}}^\top \mathbf{\Psi}^{k+1}\bm{\pi} \leqslant (\mathbf{0}^{1\times q}\;\;\mathbf{1}^{1\times (m-q)}) \mathbf{\Psi}^{k+1}\bm{\pi} = \mathbf{1}^{1\times (m-q)}\mathbf{Q}^{k+1}\bm{\nu},
        \label{SM_prob_sum2}
    \end{align}
    where $\mathbf{0}^{1\times q}$ denotes an all-zero row vector, $\mathbf{1}^{1\times (m-q)}$ denotes an all-one row vector, matrix $\mathbf{Q}$ is defined as in Eq.~\eqref{SM_psi_3}, and $\bm{\nu}\equiv(\pi(q+1),\ldots,\pi(m))^\top$ is the probability distribution in the states which are not observable Markov states. 
    
    The right hand side of Eq.~\eqref{SM_prob_sum2} indicates a process starting from the stationary distribution of $\mathbf{\Omega}$, changing state the first $q$ states (the observable Markov states) to an absorbing states, and letting the system evolve for $k+1$ steps. What we are actually evaluating is the sum of probabilities that the system does not end in the set of observable Markov states $O$, i.e.~the survival probability in the rest of the network. 
    
    We can write the generalized eigenvalues of $\mathbf{\Psi}$ as $\lambda_1 =\ldots =\lambda_q=1>|\lambda_{q+1}| \geqslant\ldots\geqslant |\lambda_{m}|$. The eigenvalues $\lambda_{q+1},\ldots,\lambda_{m}$, which are not equal to $1$, are the generalized eigenvalues of the matrix $\mathbf{Q}$. We denote the eigenvalue of $\mathbf{Q}$ with the largest modulus as $\lambda^*\equiv\lambda_{q+1}$. We show in subsection~\ref{sec:convg_bound} that the probability of remaining in a transient state decays exponentially as
    \begin{align}
        \max_{r}|(\mathbf{Q}^{k+1}\bm{\nu})_r| \leqslant M|\lambda^*|^{k+1-m^*} \binom{k+1}{m^*-1},
        \label{SM_ineq0}
    \end{align}
    where $m^*$ is the dimension of the largest Jordan block among all generalized eigenvalues. The constant $M$ is given by
    \begin{align}
        M = \max_{i}\sum_{L=1}^{R}\sum_{r=0}^{n_L-1} \sum_{l=S_L+1}^{S_L+n_L-r} \sum_{j=1}^{m-q} |u_{l-1}(i)v_{l+r-1}(j)|,
    \end{align}
    where $R$ is the total number of Jordan blocks, $n_L$ is the size of the $L$-th Jordan block $\mathbf{J}_L$, $S_L\equiv\sum_{s=0}^{L-1}n_s,\,1\leqslant L\leqslant R-1$, $S_0=1$, and $\mathbf{u}$ and $\mathbf{v}$ are generalized right and left eigenvectors of $\mathbf{Q}$. The proof is similar to proving the convergence theorem of Markov chains \cite{Rosenthal1995,Silva2016}. For completeness, we attach the proof of Eq.~\eqref{SM_ineq0} in subsection~\ref{sec:convg_bound}.
    Combining Eqs.~(\ref{SM_prob_sum}-\ref{SM_ineq0}), we have the bound 
    \begin{align}
        \sum_{z=0}^L h_k^\mathrm{N}(p=2z\delta) \leqslant \frac{1}{\pi(j)} (0\;\;\mathbf{1}^\top)\mathbf{\Psi}^{k+1}\bm{\pi} = \sum_{r=1}^{m-q}(\mathbf{Q}^{k+1}\bm{\nu})_r \leqslant C |\lambda^*|^{k+1-m^*}\binom{k+1}{m^*-1},
        \label{SM_bound1}
    \end{align}
    where we choose $C=Mm/\pi(j)$.  
        
    The bound in Eq.~\eqref{SM_bound1} vanishes as $k\to\infty$
    because $|\lambda^*|^{k}$ decays faster than the combinatorial
    grows with $k$. We can show the limit explicitly by taking the
    logarithm 
    \begin{align}
        \ln \left[|\lambda^*|^{k+1-m^*}\binom{k+1}{m^*-1} \right] = (k+1-m^*)\ln(|\lambda^*|) + \ln\left[ \frac{(k+1)!}{(m^*-1)!(k-m^*+2)!}\right].
        \label{SM_bound3}
    \end{align}
    Then we insert the bounds \cite{Robbins1955} for factorials $\sqrt{2\pi}n^{n+\frac{1}{2}}\mathrm{e}^{-n} <n! \leqslant n^{n+\frac{1}{2}}\mathrm{e}^{-n+1}$
    to bound these as 
     \begin{align}
        \ln\left[ \frac{(k+1)!}{(k-m^*+2)!}\right] 
        <& \left[\left(k+\frac{3}{2}\right)\ln(k+1)-k -\frac{1}{2}\ln(2\pi) \right. \nonumber\\
        & \left.\; +k-m^*+2- \left(k-m^*+\frac{5}{2}\right)\ln(k-m^*+2) \right] \nonumber\\
        \leqslant & (m^*-1)\ln(k+1) -m^*+2-\frac{1}{2}\ln(2\pi),
        \label{SM_bound4}
     \end{align}
    where we have inserted $m^*\geqslant 1$. Comparing the first term in Eq.~\eqref{SM_bound3} and Eq.~\eqref{SM_bound4}, the two terms depending on $k$ are $(k+1-m^*)\ln(|\lambda^*|)$ and $(m^*-1)\ln(k+1)$, and the other terms are constants. The sum of the two relevant terms goes to negative infinity when $k$ goes to infinity, because a linear function of $x$ grows faster than a logarithm. Therefore, the bound in Eq.~\eqref{SM_bound1} goes to zero when $k$ goes to infinity.
    
    Specifically, if the matrix $\mathbf{\Psi}$ is diagonalizable, we have the bound 
    \begin{align}
        \sum_{z=0}^L h_k^\mathrm{N}(p=2z\delta) = \frac{1}{\mp(j)} \!\!\sum_{S^j_k\in\{S^j_k\}^\mathrm{N}}\!\! \mp(jS^j_k) \leqslant C |\lambda^*|^{k}.
        \label{SM_bound2}
    \end{align}
    Therefore, if $\mathbf{\Psi}$ is diagonalizable, the total
    contribution from $\{S^j_k\}^\mathrm{N}$ decays as a power of
    $|\lambda^*|<1$ when $k$ becomes large. 
    
    For a fixed $k$, the contribution from $\{S^j_{k+l}\}^\mathrm{M}$
    is independent of $l$ as calculated in
    subsection~\ref{sec:convg_SkM}, and that from
    $\{S^j_{k+l}\}^\mathrm{N}$ still changes with $l$.  
    Denote the ``stationary'' histogram as $\mathbf{h}_{\infty}$, then the total variation distance (TVD) between $\mathbf{h}_{k}$ and $\mathbf{h}_{\infty}$ is bounded by the contribution from $\{S^j_{k+l}\}^\mathrm{N}$ 
         \begin{align}
        \Vert \mathbf{h}_{k} - \mathbf{h}_{\infty} \Vert_{\mathrm{TV}} &= \frac{1}{2}\sum_{z=0}^L  \left|h_k(z)-h_{\infty}(z)\right| \nonumber\\
        &= \frac{1}{2} \sum_{z=0}^L |h_k(z)- h_k^\mathrm{M}(z) - h_{\infty}(z) + h_{\infty}^\mathrm{M}(z)| \nonumber\\
        &= \frac{1}{2} \sum_{z=0}^L |h_k^\mathrm{N}(z) - h_{\infty}^\mathrm{N}(z)| \leqslant \sum_{z=0}^L h_k^\mathrm{N}(z)
        \leqslant C |\lambda^*|^{k+1-m^*}\binom{k+1}{m^*-1},
         \end{align}
    where we have written $h(p=2z\delta)$ as $h(z)$ for short and used $h_k^\mathrm{M}(z)=h_{\infty}^\mathrm{M}(z)$ in the second step.

    In conclusion, we proved that the contribution from
    $\{S^j_k\}^\mathrm{M}$ is stationary, and that
    $\{S^j_k\}^\mathrm{N}$ vanishes as $k$ tends to infinity. This
    proves that the histograms converge to a stationary distribution.

\subsection{Proof of the inequality \eqref{SM_ineq0}} \label{sec:convg_bound}
    In this subsection we will prove Eq.~\eqref{SM_ineq0}, i.e. there exists a constant $M$, such that
    \begin{align}
        \max_{r}|(\mathbf{Q}^{k+1}\bm{\nu})_r| \leqslant M|\lambda^*|^{k+1-m^*} \binom{k+1}{m^*-1}
        \label{SM_ineq0-1}
    \end{align}
    holds for any sufficiently large $k$, where $m^*$ is the dimension of the largest Jordan block among all generalized eigenvalues, $\lambda^*$ is the eigenvalue of $\mathbf{Q}$ with the largest modulus.
    
    We first express the Jordan decomposition of the matrix as $\mathbf{Q}=\mathbf{S}\mathbf{D}\mathbf{S}^{-1}$, where
    \begin{align}
        \mathbf{D} = 
        \begin{pmatrix}
            \mathbf{J}_1 & 0 & \dots & 0 \\
            0 & \mathbf{J}_2 & \ddots & \vdots \\
            \vdots & \ddots & \ddots & 0 \\
            0 & \dots & 0 & \mathbf{J}_{R}
        \end{pmatrix}, \quad
        \mathbf{J}_L = 
        \begin{pmatrix}
            \lambda_L & 1 & \dots & 0 \\
            0 & \lambda_L & \ddots & \vdots \\
            \vdots & \ddots & \ddots & 1 \\
            0 & \dots & 0 & \lambda_L
        \end{pmatrix}, \; 1\leqslant L\leqslant R.
        \label{SM_jordan}
    \end{align}
    Note that the generalized eigenvalues are enumerated such that each $\lambda_L$ corresponds to a Jordan block with $g$ repeated eigenvalues $\lambda_{L_1}=\ldots=\lambda_{L_g}$ and $g$ is the geometric multiplicity. $R$ is the total number of Jordan blocks. 
    Columns of $\mathbf{S}$ are generalized right eigenvectors
    $\mathbf{u}$ and rows of $\mathbf{S}^{-1}$ are generalized left
    eigenvectors $\mathbf{v}$ of $\mathbf{Q}$.  
    If $\mathbf{Q}$ is diagonalizable, then $\mathbf{D}$ is a diagonal matrix, i.e.~$R=m-q$.

    The $k$-th power of matrix $\mathbf{Q}$ can be written as
     \begin{align}
        \mathbf{Q}^{k} &= \mathbf{S}\mathbf{D}^k\mathbf{S}^{-1} \nonumber\\
        &=
        \begingroup 
        \setlength\arraycolsep{2pt}
        \begin{pmatrix}
            u_1(1) & \dots & u_{m-q}(1) \\
            u_1(2) & \dots & u_{m-q}(2) \\
            \vdots & \ddots & \vdots \\
            u_1(m-q) & \dots & u_{m-q}(m-q)
        \end{pmatrix}
        \begin{pmatrix}
            \mathbf{J}_1^k & 0 & \dots & 0 \\
            0 & \mathbf{J}_2^k & \ddots & \vdots \\
            \vdots & \ddots & \ddots & 0 \\
            0 & \dots & 0 & \mathbf{J}_{R}^k
        \end{pmatrix}
        \begin{pmatrix}
            v_1(1) & \dots & v_1(m-q) \\
            \vdots & \ddots & \vdots \\
            v_{m-q}(1) & \dots & v_{m-q}(m-q)
        \end{pmatrix}.
        \endgroup
        \label{thm4_mat1}
     \end{align}
    Let $n_L$ be the size of the Jordan block $\mathbf{J}_L$. For $k\geqslant n_L-1$, we have
    \begin{align}
        \mathbf{J}_L^k = 
        \begin{pmatrix}
            \lambda_L^k & \binom{k}{1}\lambda_L^{k-1} & \dots & \dots & \binom{k}{n_L-1}\lambda_L^{k-n_L+1} \\
            0 & \lambda_L^k & \binom{k}{1}\lambda_L^{k-1} & \dots & \binom{k}{n_L-2}\lambda_L^{k-n_L+2} \\
            \vdots & 0 & \lambda_L^k & \ddots & \vdots \\
            \vdots & \vdots & \ddots & \ddots & \binom{k}{1}\lambda_L^{k-1} \\
             0 & 0 & \dots & 0 & \lambda_L^k
        \end{pmatrix}.
    \end{align}
    Define $S_0=1,$ and $S_L=\sum_{s=0}^{L-1}n_s,\,1\leqslant L\leqslant R-1$. We can explicitly express the $i$-th element of $\mathbf{Q}^{k}\bm{\nu}$ as
     \begin{align}
        (\mathbf{Q}^{k}\bm{\nu})_{i} &= \sum_{l,t=1}^{m}\sum_{j=1}^{m-q} (\mathbf{S})_{il} (\mathbf{D}^k)_{lt} (\mathbf{S}^{-1})_{tj}\nu(j) \nonumber\\
        &= \sum_{L=1}^{R}\,\sum_{l,t=S_L+1}^{S_L+n_L} \sum_{j=1}^{m-q} (\mathbf{S})_{il} (\mathbf{D}^k)_{lt} (\mathbf{S}^{-1})_{tj} \nu(j) \nonumber\\
        &= \sum_{L=1}^{R} \sum_{r=0}^{n_L-1}\binom{k}{r} \lambda_L^{k-r} \sum_{l=S_L+1}^{S_L+n_L-r} \sum_{j=1}^{m-q} u_{l-1}(i)v_{l+r-1}(j) \nu(j).
     \end{align}
    Define $m^*$ as the size of the largest Jordan block. For $k>2(m^*-1)$ we have
     \begin{align}
        |(\mathbf{Q}^{k}\bm{\nu})_{i}| &\leqslant 
        \sum_{L=1}^{R-1} \sum_{r=0}^{n_L-1}\binom{k}{r} |\lambda_L|^{k-r} \sum_{l=S_L+1}^{S_L+n_L-r} \sum_{j=1}^{m-q} |u_{l-1}(i)v_{l+r-1}(j)\nu(j)| \nonumber\\
        &\leqslant \binom{k}{m^*-1} |\lambda^*|^{k-m^*} \sum_{L=1}^{R-1}\sum_{r=0}^{n_L-1} \sum_{l=S_L+1}^{S_L+n_L-r} \sum_{j=1}^{m-q} |u_{l-1}(i)v_{l+r-1}(j)|,
     \end{align}
    where we inserted $0\leqslant\nu(j)\leqslant 1$ in the last step.
    Define a constant $M$ which does not depend on $k$ as
    \begin{align}
        M = \max_{i}\sum_{L=1}^{R}\sum_{r=0}^{n_L-1} \sum_{l=S_L+1}^{S_L+n_L-r} \sum_{j=1}^{m-q} |u_{l-1}(i)v_{l+r-1}(j)|.
    \end{align}
    Therefore, we have proved that for any $k>2(m^*-1)$, for any initial distribution $\bm{\mu}_0$,
    \begin{align}
        \max_{i}|(\mathbf{Q}^{k}\bm{\nu})_{i}| \leqslant M|\lambda^*|^{k-m^*} \binom{k}{m^*-1} .
    \end{align}

\section*{Appendix C:~Conditional probability for the second step in the future} \label{sec:cond_prob_two_step}
\renewcommand{\theequation}{C.\arabic{equation}}
\setcounter{equation}{0}
\renewcommand{\thefigure}{C\arabic{figure}}
\setcounter{figure}{0}

We first state a sufficient condition where for some $j$ and $i$, for any $k$, $\mp(i|jS^j_k)$ is a delta distribution, i.e.\ Eq.~\eqref{def_split_prob1} is independent of $S^j_k$:
$\mp(i|\aa_{k+1})$ are the same for any $\aa_{k+1}$ in $j$. An example for the dynamics for reduced sequences satisfying this condition, which yields $\mp(i|jS^j_k)=p$ independent of the history, is shown in Fig.~(\ref{fig_Appx_2}). Another example satisfying the condition can be found in \cite{Igoshin2025}.  

Now suppose that for some observable state $j$, for any $i$ directly
connected to $j$ and any $k$, all the histograms show delta
distributions at the same location $\mp(i|j)$. The transition from $j$
to one step in the future does not display memory. In this case, state
$j$ may be, but not necessarily is, a Markov state. Nonetheless, we can
check the conditional probability of arriving at state $o$ at the
$(k+3)$-th step conditioned on being at $j$ at the $(k+1)$-th step
with a history $S_k$, as   
\begin{align}
     \mp(s_{k+3}=o|s_{k+1}=j;S_k) 
    = \frac{\sum_{\aa_{k+1}}\!\! \ind{j}[\aa_{k+1}] \mp(\aa_{k+1}S^j_k) \sum_{\aa_{k+2}}\!\! \mp(o|\aa_{k+2})\mp(\aa_{k+2}|\aa_{k+1})}{\sum_{\aa_{k+1}} \ind{j}[\aa_{k+1}] \mp(\aa_{k+1}S^j_k)}, 
    \label{appdx_cond_prob}
\end{align}
where $\mp(o|\aa_{k+2}) = \sum_{\aa_{k+3}}\!\! \ind{o}[\aa_{k+3}]
\mp(\aa_{k+3}|\aa_{k+2})$. 
Note that $o$ is a second-order neighbor of $j$ in the connectivity
graph. If the histograms for $\mp(s_{k+3}=o|s_{k+1}=j;S_k)$ depend on
$S_k$, i.e.\ are \emph{not} delta distributions at the same location,
then state $j$ is in fact non-Markov.  
\begin{figure}[h]
    \centering
    \includegraphics[width=0.35\textwidth]{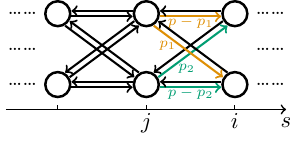}
    \caption{A part of a network for the dynamics of the reduced microscopic sequence. The values on arrows represent elements in the transition matrix $\mathbf{\Omega}$ for the reduced sequence. The two microscopic states in the observable state $j$ both satisfy $\mp(i|\aa_{k+1})=p$. }
    \label{fig_Appx_2}
\end{figure}

If some histograms for $\mp(s_{k+3}=o|s_{k+1}=j;S_k)$ are not delta distributions, we can also tell, from Eq.~\eqref{appdx_cond_prob}, that the sums $ \sum_{\aa_{k+2}}\!\! \mp(o|\aa_{k+2})\mp(\aa_{k+2}|\aa_{k+1}) = \mp(s_{k+3}=o|\aa_{k+1})$ are not the same for all the $\aa_{k+1}$ in state $j$. The
contrapositive statement can easily be shown by assuming the sum $\mp(s_{k+3}=o|\aa_{k+1})$ is a constant for any $\aa_{k+1}$. Then, the summation over $\aa_{k+1}$ in Eq.~\eqref{appdx_cond_prob} cancels.  
Furthermore, the dependence of $\mp(s_{k+3}=o|\aa_{k+1})$ on $\aa_{k+1}$ means that $\mp(o|\aa_{k+2})$ are not the same for all the $\aa_{k+2}$ directly connected to any microscopic states in
$j$. Otherwise, we have that $\mp(s_{k+3}=o|\aa_{k+1})= \mp(o|\aa_{k+2}) \sum_{\aa_{k+2}}\!\!\mp(\aa_{k+2}|\aa_{k+1}) = \mp(o|\aa_{k+2})$ does \emph{not} depend on $\aa_{k+1}$, which
contradicts the first statement. 
Note that all the statements here are uni-directional. The statements above show that investigating memory locally yields information on microscopic splitting probabilities, even though memory is determined by the entire network.

Similarly, we can always check for any step further in the future $\mp(s_{k+\Delta}=o|s_{k+1}=j;S_k)$ ($\Delta\geqslant 2$) to be more ``certain'' about state $j$ being a Markov state, while it is generally \emph{not} practically feasibly (for obvious reasons) to check the probabilities of arbitrary future sequences conditioned on histories. 

Additionally, if a bar at some location $p$ disappears as $k$ increases (e.g.\ the bar at $p=0.8$ in Fig.~\ref{fig1}(c-d)), then  it arises from history sequences which do \emph{not} visit an observable Markov state. This can also be used to detect a non-Markov state in the case where for some $j$ and $i$, for any $k$, $\mp(i|jS^j_k)$ is a delta distribution.

\section*{Appendix D:~Detection of observable Markov state} \label{sec:detection}
\renewcommand{\theequation}{D.\arabic{equation}}
\setcounter{equation}{0}

In this appendix, we explore the possibility to detect practically an observable Markov state, which is a type of a ``Markovian event'' \cite{Meer2023,Deguenther2024,Deguenther2024a}. We will show that for an $n$-th order Markov process, we only need to check the Markov property up to $n$ steps in the future and $n$ steps in the past.

In the following calculation, we write $\{s_{t}\}_{t_1\leqslant t\leqslant t_2}$ as $\{s\}_{t_2,t_1}$ for short. Note that the time points $t_2,\,t_1$ are written in a reverse order.
We first recap the definition of a Markov state: for any $k>g\geqslant 1$, for any future sequence $\{s\}_{k,g+1}$ and history sequence $\{s\}_{g-1,1}$, the observable Markov state $\gamma$ satisfies
\begin{align}
    \mp(\{s\}_{g,k+2}|s_{k+1}\in O;\{s\}_{k,1}) = \mp(\{s_{t}\}_{g,k+2}|s_{k+1}\in O) .
    \label{def_Markov_state1}
\end{align}
For $g=k+2$ (one step in the future) we can test with the method of histograms shown in the main text. We further discuss longer future sequences where $g\geqslant k+3$.
From the definition of conditional probability, we have the identity
\begin{align}
    \mp(s_{g}|\{s\}_{g-1,k+2};s_{k+1}\in O;\{s\}_{k,1}) &= \frac{\mp(\{s\}_{g,k+2};s_{k+1}\in O;\{s\}_{k,1})}{\mp(\{s\}_{g-1,k+1};s_{k+1}\in O;\{s\}_{k,1})} \nonumber\\ 
    &= \frac{\mp(\{s\}_{g,k+2}|s_{k+1}\in O;\{s\}_{k,1})}{\mp(\{s\}_{g-1,k+2}|s_{k+1}\in O;\{s\}_{k,1})}.
    \label{SM_cond1}
\end{align}
Now we \emph{assume} that the observed dynamics is an $n$-th (definite) order Markov process. When the length of the sequence in the future satisfies $g-k-1\geqslant n$, i.e. $g-n\geqslant k+1$, then for any $s_g$ we can truncate the sequence in the condition from $\{s\}_{g-1,1}$ to $\{s\}_{g-1,g-n}$, and extend it to $\{s\}_{g-1,k+1}$ as
\begin{align}
    \mp(s_g|\{s\}_{g-1,k+2};s_{k+1}\in O;\{s\}_{k,1}) = \mp(s_g|\{s\}_{g-1,g-n}) = \mp(s_g|\{s\}_{g-1,k+2};s_{k+1}\in O).
    \label{nth_order}
\end{align}
Given the condition in Eq.~\eqref{nth_order}, combining the identity in Eq.~\eqref{SM_cond1}, we have the statement
\begin{align}
    & \mp(\{s\}_{g-1,k+2}|s_{k+1}\in O;\{s\}_{k,1}) = \mp(\{s\}_{g-1,k+2}|s_{k+1}\in O) \nonumber\\ 
    \Rightarrow \; & \mp(\{s\}_{g,k+2}|s_{k+1}\in O;\{s\}_{k,1}) = \mp(\{s\}_{g,k+2}|s_{k+1}\in O).
    \label{SM_cond2}
\end{align}
Note that the right-hand side of Eq.~\eqref{SM_cond2} is the condition to test an observable Markov state. Thus, for an $n$-th order Markov process, we only need to check the condition in Eq.~\eqref{def_Markov_state1} for $k\leqslant n$.

\section*{Appendix E:~Simulation details} \label{sec:sim}
\renewcommand{\theequation}{E.\arabic{equation}}
\setcounter{equation}{0}
For the 8-state toy protein model in Figure~\ref{fig1}(a) of the main text, we
randomly generate free energies $F_i,\,0\leqslant F_i\leqslant 10$ for
each state $i$. The transition rates are then determined by $L_{ji}=\exp((F_i-F_j)/2)$ for any $i\neq j$. The transition rate matrix $\mathbf{L}$ for the example in the main text is
\begin{align}
    \mathbf{L}=
    \begin{pmatrix}
        -2.543   & 0.448 & 4.115 & 15.231 & 0 &   0 &   0 &   0  \\
        2.234 & -0.680 &   0 &   0 &   7.416 & 10.275 & 0 &   0 \\  
        0.243 & 0 &   -14.864 &   0 &   0.806 & 0 &   0.105 & 0 \\  
        0.066 & 0 &   0 &   -53.733 &   0 &   0.302 & 0.028 & 0  \\ 
        0 &   0.135 & 1.240 & 0 &   -8.937 &   0 &   0 &   1.399 \\
        0 &   0.097 & 0 &   3.312 & 0 &   -11.568 &   0 &   1.010 \\
        0 &   0 &   9.509 & 35.190 & 0 &   0 &   -0.227 &  10.729 \\
        0 &   0 &   0 &   0 &   0.715 & 0.990 & 0.093 & -13.138 \\  
    \end{pmatrix}.
\end{align}
The corresponding transition matrix $\mathbf{\Phi}$ for the microscopic sequence, i.e. the splitting probability matrix, is given by
\begin{align}
    \mathbf{\Phi} =
    \begin{pmatrix}
        0 &  0.658 & 0.277 & 0.283 & 0 &  0 &  0 &  0  \\
        0.879 & 0 &  0 &  0 &  0.830 & 0.888 & 0 &  0   \\
        0.096 & 0 &  0 &  0 &  0.090 & 0 &  0.464 & 0   \\
        0.026 & 0 &  0 &  0 &  0 &  0.026 & 0.125 & 0   \\
        0 &  0.198 & 0.083 & 0 &  0 &  0 &  0 &  0.106 \\
        0 &  0.143 & 0 &  0.062 & 0 &  0 &  0 &  0.077 \\
        0 &  0 &  0.640 & 0.655 & 0 &  0 &  0 &  0.817 \\
        0 &  0 &  0 &  0 &  0.080 & 0.086 & 0.411 & 0   
    \end{pmatrix}.
\end{align}
We simulate the discrete-time Markov chain to generate one microscopic
trajectory with $10^8$ steps, which is equivalent to simulating
the continuous-time process and recording only the visited states. Then we project the trajectory onto observable states, and remove consecutive steps in the same observable state to obtain the observed sequence. We use this observed sequence for further analysis based on histograms.

\section*{Appendix F:~Examples of driven systems} \label{sec:driven}
\renewcommand{\theequation}{F.\arabic{equation}}
\setcounter{equation}{0}
\renewcommand{\thefigure}{F\arabic{figure}}
\setcounter{figure}{0}

The histogram method introduced in the main text applies also to
driven systems, while it does not provide any additional information on the irreversibility of the system. In this section we show an example with the same connection as the network in the main text but different transition rates such that detailed balance is broken.

We drive the system by adding a matrix $\mathbf{\Phi}^A$ to the equilibrium transition matrix $\mathbf{\Phi}$. We construct $\mathbf{\Phi}^A$ to be anti-symmetric with respect to the invariant measure (steady state distribution) \cite{Kaiser2017}, so that the steady state of the dynamics of the microscopic sequence does not change. Here we choose to drive the system along states $1-2-5-3-1$ by assigning the nonzero elements in the matrix $\mathbf{\Phi}^A$ to be
\begin{align}
    & \Phi^A_{21}=\varphi_1,\, \Phi^A_{52}=\varphi_2,\, \Phi^A_{35}=\varphi_5,\,\Phi^A_{13}=\varphi_3,\, \nonumber\\
    & \Phi^A_{12}=-\varphi_2,\, \Phi^A_{25}=-\varphi_5,\, \Phi^A_{53}=-\varphi_3,\, \Phi^A_{31}=-\varphi_1,
\end{align}
where we define $\varphi_i \equiv \varphi/\pi_i$, and $\pi_i$ is the steady state distribution of the Markov chain $\mathbf{\Phi}$ at state $i$. The constant $\varphi$ is a parameter that controls the strength of driving. Histograms of transition probabilities between different pairs of states conditioned on history of different lengths $k$ for driving parameter $\varphi=0.05$ and $\varphi=-0.05$ are displayed in Figure~\ref{fig_SM_2}. The analysis and obtained information from the driven systems are basically the same as the equilibrium system shown in the main text.

\begin{figure}[h]
    \centering
    \includegraphics[width=1.0\textwidth]{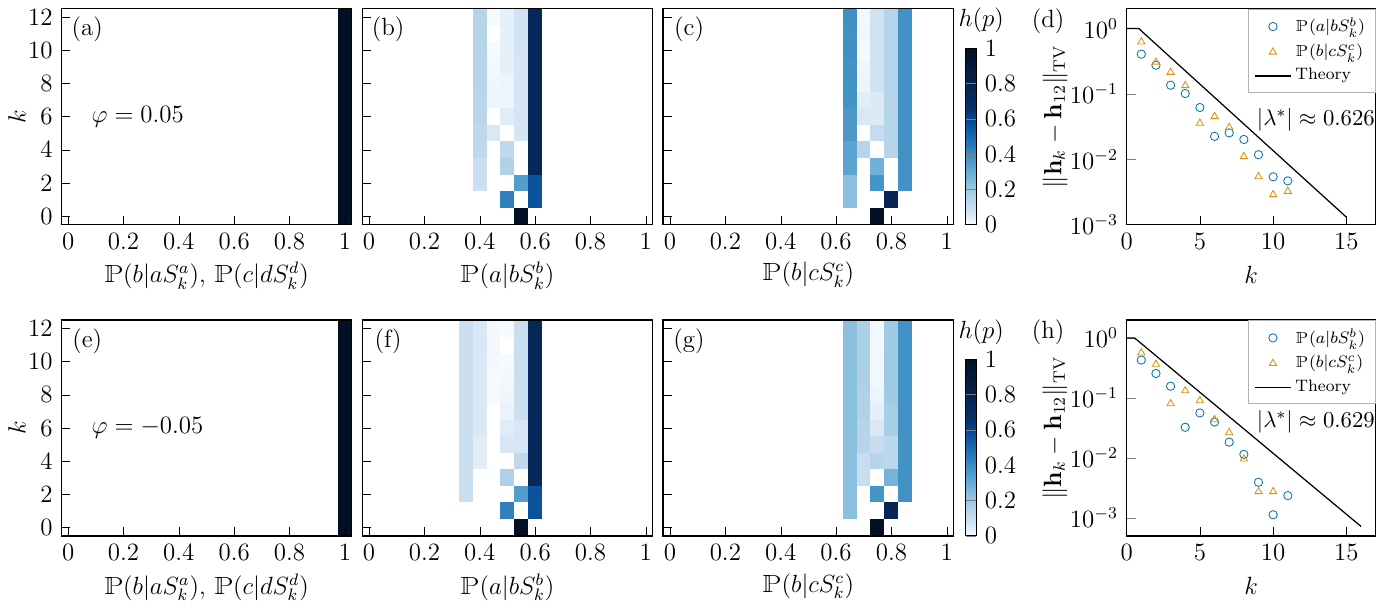}
    \caption{(a)-(c)~Histograms of transition probabilities between different pairs of states conditioned on history of different lengths $k$ for driving parameter $\varphi=0.05$. Each row represents a histogram for a fixed history length $k$. The bar width in each histogram is $2\delta=0.05$. The color bar is the same for all plots. (d) The points show the TVD determined from simulation results, and the solid line is the theoretical bound which is independent of the pair of states. (e)-(g)~Histograms of transition probabilities for driving parameter $\varphi=-0.05$. (h) The TVD determined from simulation results and the theoretical bound. 
    }
    \label{fig_SM_2}
\end{figure}

\section*{Appendix G:~Discussion on insufficient statistics} \label{sec:sampling}
\renewcommand{\theequation}{G.\arabic{equation}}
\setcounter{equation}{0}
\renewcommand{\thefigure}{G\arabic{figure}}
\setcounter{figure}{0}

We generate from simulation a trajectory with $T=10^8$ transitions for sufficient statistics of history sequences up to $k_{\mathrm{max}}=12$ steps. If the statistics is not sufficient, i.e.\ with to short
trajectories, the histograms for large $k$ will scatter because the transition probability $\mp(a|bS^b_k)$ for each specific $S^b_k$ is \emph{not} a delta peak but a distribution around the true value, as shown in Figure~\ref{fig_SM_3}.  The changes of histograms due to memory (the rows at $2\leqslant k\leqslant 6$ in Figure~\ref{fig_SM_3}(a)) and those due to undersampling (the rows at $k\geqslant 5$ in Figure~\ref{fig_SM_3}(d)) display \emph{different} characteristics. The former typically involves the disappearance and emergence of some bars every time $k$ increases by $1$. In contrast, the later shows a scatter around some true values, and there are only new bars emerging but no bars completely vanish.

\begin{figure}[h]
    \includegraphics[width=1.0\textwidth]{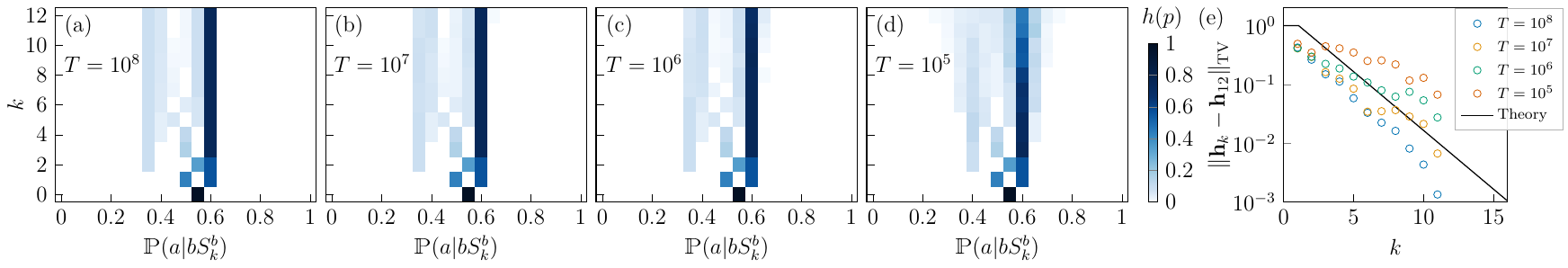}
    \caption{(a)-(d)~Histograms of transition probabilities $\mp(a|bS^b_k)$ conditioned on history of different lengths $k$ for different trajectory length. $T$ denotes the number of transitions in the trajectory. (e) The points show the TVD determined from simulation results, and the solid line is the theoretical bound.
    }
    \label{fig_SM_3}
\end{figure}

\section*{Appendix H:~Example of a network without observable Markov states} \label{sec:noMarkov}
\renewcommand{\theequation}{H.\arabic{equation}}
\setcounter{equation}{0}
\renewcommand{\thefigure}{H\arabic{figure}}
\setcounter{figure}{0}
In this appendix, we present an example of a network without an observable Markov state, shown in Figure~\ref{fig_SM_4}. The network consists of 6 microscopic states projected onto 3 observable states $a=\{1,2\}$, $b=\{3,4\}$, and $c=\{5,6\}$. The transition rates are randomly drawn in the same way as in Appendix~\hyperref[sec:sim]{E}. The simulated trajectory of the Markov chain consists of $10^8$ steps. The transition matrix $\mathbf{\Phi}$ in this example is given by
\begin{align*}
    \mathbf{\Phi}=
    \begin{pmatrix}
        0   & 0.126 & 0.096 & 0.232 & 0 &   0  \\
        0.205 &   0 & 0.171 & 0.415 & 0 &   0 \\  
        0.046 & 0.051 &   0 & 0.093 & 0.053 & 0.055 \\  
        0.749 & 0.823 & 0.625 & 0   & 0.859 & 0.885  \\ 
        0 &   0 & 0.043 & 0.104 & 0   & 0.061 \\
        0 &   0 & 0.064 & 0.156 & 0.088 & 0 \\  
    \end{pmatrix},
\end{align*}
where values are rounded to 3 decimal places.

As shown in Figure~\ref{fig_SM_4}, the difference between the histogram at $k$ and $k_{\rm max}$ decays, indicating the convergence of histograms, even though the theoretical bound Eq.~\eqref{bound} does not apply in this system.

\begin{figure}[h]
    \centering
    \includegraphics[width=0.75\textwidth]{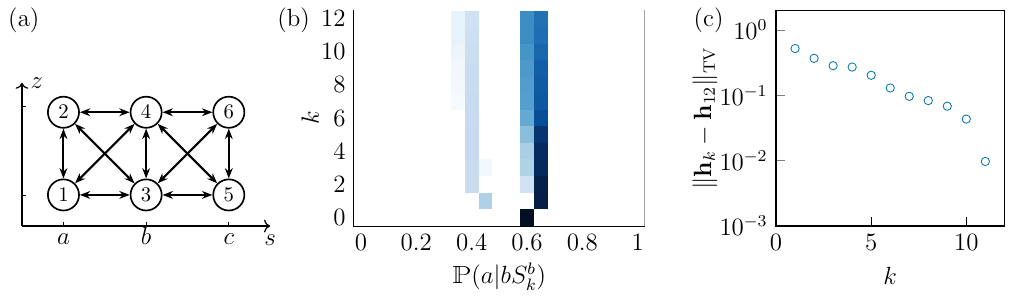}
    \caption{(a)~Schematic of the full Markov network. THe $s$-axis is the observable and $z$-axis the hidden dimension. (b)~Histograms of transition probabilities $\mp(a|bS^b_k)$ conditioned on history of different lengths $k$. (c)~The TVD determined from simulation results.
    }
    \label{fig_SM_4}
\end{figure}

\end{appdxsection}

\FloatBarrier
\clearpage

\section*{References}
\bibliography{manuscript_memory_histogram}

\end{document}